\documentclass[twocolumn,10pt]{IEEEtran}
\usepackage{ifpdf, flushend,subfigure}

\ifCLASSINFOpdf
  \usepackage[pdftex]{graphicx}
	
  \graphicspath{{../pdf/}{../jpeg/}}
  \DeclareGraphicsExtensions{.pdf,.jpeg,.png}
\else
  \usepackage[dvips]{graphicx}
  \graphicspath{{../eps/}}
  \DeclareGraphicsExtensions{.eps}
\fi
\usepackage[cmex10]{amsmath}
\usepackage {amssymb}
\usepackage{algorithmic}
\usepackage{array}
\usepackage{mdwmath}
\usepackage{mdwtab}
\usepackage{eqparbox}
\usepackage{url}
\usepackage{hyperref}
\usepackage{algorithm}
\usepackage{algorithmic}

\newcommand{\argmin}{\operatornamewithlimits{argmin}}

\newcommand{\beq}{\begin{equation}}
\newcommand{\eeq}{\end{equation}}
\newcommand{\beqn}{\begin{eqnarray}}
\newcommand{\eeqn}{\end{eqnarray}}
\newcommand{\beqno}{\begin{eqnarray*}}
\newcommand{\eeqno}{\end{eqnarray*}}
\newcommand{\bma}{\begin{displaymath}}
\newcommand{\ema}{\end{displaymath}}
\newcommand{\bnu}{\begin{enumerate}}
\newcommand{\enu}{\end{enumerate}}
\newcommand{\bce}{\begin{center}}
\newcommand{\ece}{\end{center}}
\newcommand{\btb}{\begin{tabular}}
\newcommand{\etb}{\end{tabular}}

\begin{document}

%
\title{Joint Data Compression and MAC Protocol Design for Smartgrids with Renewable Energy}

\author{Le~Thanh~Tan,~\IEEEmembership{Member,~IEEE,} Long~Bao~Le,~\IEEEmembership{Senior Member,~IEEE} 
\thanks{ Manuscript received December 23, 2014; accepted June 11, 2016. 
The editor coordinating the review of this paper and approving it for publication is  Dr. Yun Rui. 
}
\thanks{L. T. Tan is with the School of Electrical Computer and Energy Engineering, Arizona State University (ASU), Tempe, AZ, USA. E-mail: tlethanh@asu.edu. }
\thanks{L. B. Le is with Institut  National  de  la  Recherche  Scientifique--
\'{E}nergie, Mat\'{e}riaux et T\'{e}l\'{e}communications, Universit\'{e} du Québec, Montr\'{e}al,
QC J3X 1S2, Canada. E-mail: long.le@emt.inrs.ca. } 
}


\markboth{Wireless Communications and Mobile Computing} {Le \MakeLowercase{\textit{et al.}}: Joint Data Compression and MAC Protocol Design for Smartgrids with Renewable Energy}
\markright{Wireless Communications and Mobile Computing} 

\maketitle

\begin{abstract}
\boldmath
In this paper, we consider the joint design of data compression and 802.15.4-based medium access control (MAC) protocol for smartgrids with renewable energy.
We study the setting where a number of nodes, each of which comprises electricity load and/or renewable sources, report periodically their injected powers to 
a data concentrator. Our design exploits the correlation of the reported data in both time and space
to efficiently design the data compression using the compressed sensing (CS) technique and 
the MAC protocol so that the reported data can be recovered reliably within minimum reporting time. Specifically,
we perform the following design tasks: i) we employ the two-dimensional (2D) CS technique to compress
the reported data in the distributed manner; ii) we propose to adapt the 802.15.4 MAC protocol frame structure
to enable efficient data transmission and reliable data reconstruction; and iii) we develop an analytical model
based on which we can obtain efficient MAC parameter configuration to minimize the reporting delay.
Finally, numerical results are presented to demonstrate the effectiveness of our proposed framework compared to existing solutions.
\end{abstract}

\begin{IEEEkeywords}
CSMA MAC protocols, renewable energy, compressed sensing, smart grids, and power line communications.
\end{IEEEkeywords}
\IEEEpeerreviewmaketitle

\section{Introduction}
\label{Intro}

The future energy grid is expected to integrate more distributed and renewable
energy resources with significantly enhanced communications infrastructure 
for timely and reliable data exchanges between the control center and various grid control and monitoring points \cite{Stojmenovic14}.
Smartgrid is an example of the cyber-physical system (CPS) that integrates different communications, control, and computing
technologies \cite{Zanella14}. Smartgrid communications infrastructure is an important component of the future smartgrid that 
enables to support many critical grid control, monitoring, and management
operations and emerging smartgrid applications \cite{Zanella14}--\cite{Wang14}. The smartgrid communications infrastructure is typically
hierarchical, i.e., data communications between customer premises (smart meters (SMs)) and local concentrators and 
between local concentrators and the utility company are supported by field/neighborhood area networks and long-haul wide area networks, respectively \cite{Sendin12}--\cite{Botte05}. 
The former is usually based on the low bandwidth communications technologies such as Zigbee, WiFi, and power line communications (PLC) while 
the later is required to have higher capacity, which can be realized by employing LTE, 3G cellular, WiMAX, and fiber optics for example.

Our current work concerns the design of data compression and MAC protocol for the field/neighborhood area network where PLC is employed to report 
injected powers from grid reporting points to the local concentrator. 
In fact, several smartgrid projects in Spain \cite{Sendin12}, and France \cite{ERDF} have chosen PLC for smartgrid deployment since
PLC can be realized with low-cost modems and it can utilize available electricity wires for data communications.
In addition, as reported in Italy's Telegestore project, SMs have been installed at customer premises 
to send data to a concentrator via PLC \cite{Botte05}.
We focus on reporting injected powers at different grid reporting points once in every reporting interval (RI) \cite{Alam13, Abdel13}. This is motivated
by the fact that information on injected powers can be very useful for various grid applications
such as line failure prediction \cite{Rao11}--\cite{Sendin13} or congestion management and grid control applications \cite{Lo12, Lo13}.
Furthermore, the utility control center can utilize the collected data to further estimate the complete phasor data at different nodes \cite{Abur04}  which can be then used
in the control of voltages and reactive powers or in the active load management \cite{Alam13, Abdel13, Samarakoon11}, outage management \cite{Sridharan01, Liu02}
as well as illegal electricity usages and data intrusion/anomaly detection \cite{Cavdar04, Valenzuela13}.

There have been some existing works that study data compression and MAC protocol design issues in both smartgrids and wireless network contexts.
There are two popular standards for the PLC technology, namely PRIME and G3-PLC,
whose physical- and MAC-layer design aspects are investigated via simulations in \cite{Matan13a}--\cite{Patti13}. In addition,
 the authors in \cite{Alam13} study the state estimation problem where the voltage phasors at different nodes are recovered based on limited
collected data. Li et al. \cite{Hush10} propose to employ the CS technique for sparse electricity data compression.
The authors in \cite{Fazel11} consider random access exploiting the CS capability for energy efficiency communications in wireless sensor networks.
However, none of these existing works considers the joint design of communications access and data compression that supports the communication for 
a large number of nodes (e.g., in smartgrid communication infrastructure). We aim to fill this gap in the current work. In particular, we make the following contributions.

\begin{itemize}

\item We propose the CS-based data compression and PRIME MAC protocol design that supports the communication between
a number of grid reporting points and a data concentrator using PLC and random access. 
Specifically, we consider a distributed random reporting (DRR) mechanism where each node decides to report its data in the probabilistic manner 
by using the 802.15.4-based MAC protocol. Then the Kronecker CS technique (2D CS), which can exploit the spatio-temporal correlation of data, is employed for
 data reconstruction at the control center. 

\item We develop an analytical model for the MAC protocol, which enables us to achieve efficient and reliable data reconstruction using compressed sensing. 
In addition, we present an algorithm which determines efficient configuration for MAC parameters so that the average reporting time is minimized. 
Also, we analyze the energy consumption and bandwidth usage for our proposed design.

\item We present numerical results to demonstrate the significant performance gains of our proposed design compared to the 
non-compressed solution and the centralized time division multiple access (TDMA) scheme. These performance gains are demonstrated
for reporting delay, energy consumption, and bandwidth usage.

\end{itemize}

The remainder of this paper is organized as follows.
Section \ref{SystemModel} presents the system model and Section \ref{GCSMAC} describes a CS-based data compression.
Section \ref{Net_des_Per_Ana} describes our design and performance analysis. Numerical results are demonstrated in Section \ref{Results} followed by
the conclusion in Section \ref{Conclusion}.
 
\section{System Model}
\label{SystemModel}

\begin{figure*}[!t]
\centering
\includegraphics[width=160mm, height=120mm]{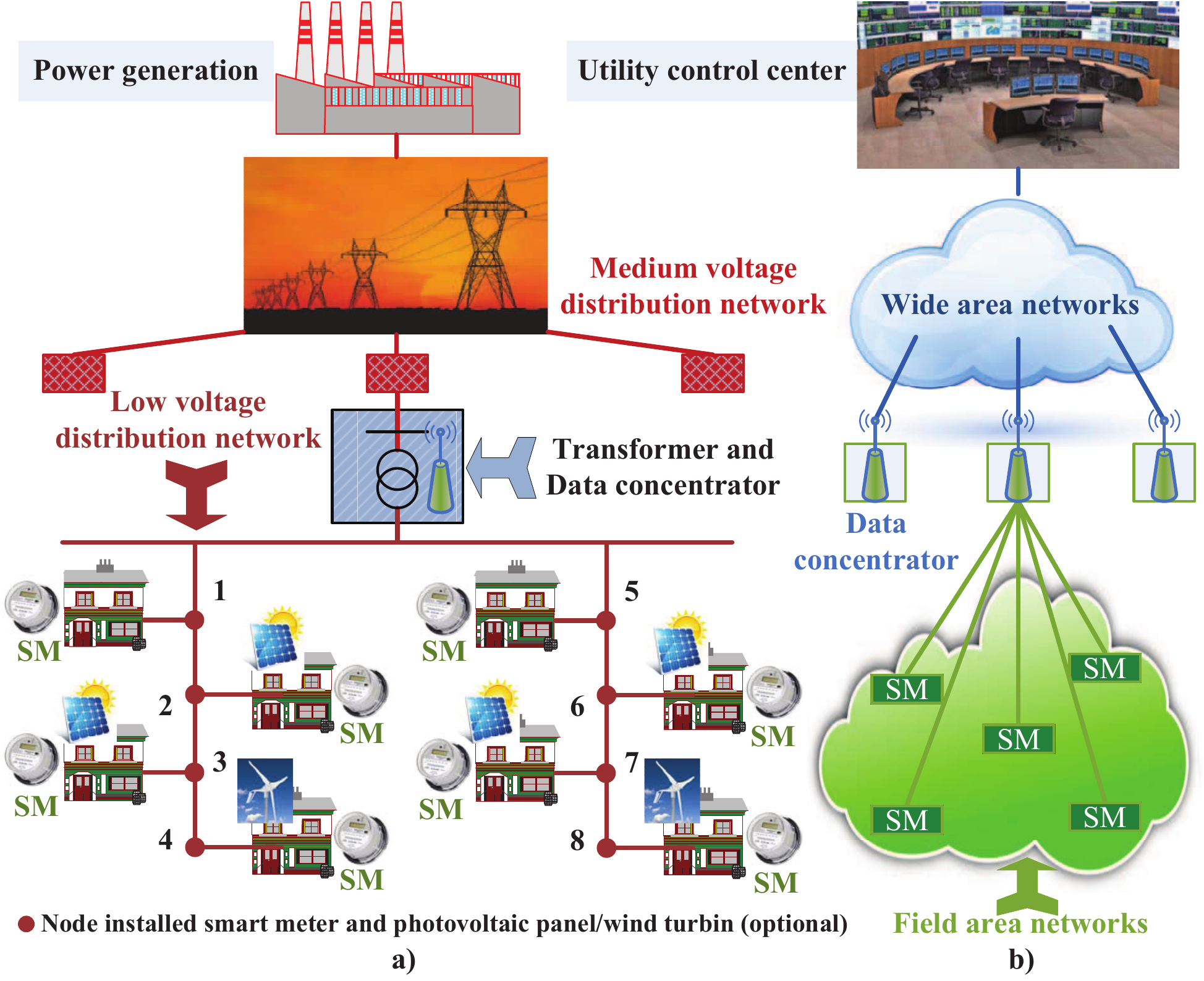} 
\caption{Next-generation smartgrid with a) power grid; b) communications infrastructure.}
\label{LV_network}
\end{figure*}

We consider the communications between $n_S$ grid reporting points and one data concentrator
which corresponds to the point-to-multipoint communication in the field
area network illustrated in Fig.~\ref{LV_network}. There would be
a large number of data concentrators collecting various types of monitoring and control data in the distribution network; however, we focus on 
the design of the point-to-multipoint communication for one such data concentrator without loss
of generality. For brevity, we refer to the grid reporting points as nodes 
in the following. Each node is assumed to comprise a load and/or a solar/wind generator.
Deployment of such renewable generators in large scale at customer 
premises, light poles, or advertising panels has been realized for several urban areas in Europe \cite{Sendin12}--\cite{Botte05}.

We further assume that each node is equipped with a SM, which is capable of reporting
power data to the data concentrator using the PLC. Upon reaching the data concentrator,
the collected data is forwarded to the utility control center via the high-speed backhaul network,
which is assumed to be error free.\footnote{This assumption is reasonable since the backhaul network
typically has very high capacity and reliability. This is the case if the backhaul network is realized by
fiber optics for example.} The data concentrator can be installed at the same place as the transformer to collect power data from its nodes 
as shown in Fig.~\ref{LV_network}, which enables reliable communications. Note, however, that data concentrator can be deployed
in the medium voltage (MV) line as well where the power data is transferred over medium-voltage power lines passing coupling devices at transformers \cite{Sendin11}.

We assume that each node must report the injected power value once for each fixed-length RI \cite{Alam13, Abdel13}.
Such collected injected powers can be used for various applications as discussed in Section~\ref{Intro}.
The RI typically ranges from 5 minutes to one day, which is configurable for most available SMs \cite{SMII} even though
we set RI equal to 5 minutes in this work.

Let $\mathbf{S}_{l,i}$, $\mathbf{S}_{g,i}$ and $\mathbf{S}_i$ be the load, distributed generation and injected powers at node $i$, respectively.
Then, the injected power at node $i$ can be expressed as $\mathbf{S}_i = \mathbf{S}_{g,i} - \mathbf{S}_{l,i}$.
We assume that the control center wishes to acquire information about $\mathbf{S}_i$ for all nodes $i$ $(i \in \left[1, n_S\right])$
in each RI.\footnote{The proposed framework can be applied to other types of grid data as long as they exhibit sparsity in the
time and/or space domains.} Due to the low bandwidth constraint of PLC 
\cite{Matan13a}--\cite{Patti13}, our objectives are to perform joint design of data compression 
and MAC protocol so that reliable data reporting can be achieved within minimum reporting time (RT). 
This is motivated by the fact that most smartgrid control and monitoring applications have very stringent
delay  requirements. Another important design target is that
the MAC protocol is distributed so that it can allow low-cost and large-scale deployment.

It has been shown in some recent works that power grid data typically exhibits
strong correlation over space and time \cite{Hu11}--\cite{Chen10b}. Therefore, the CS technique can be employed
for data compression, which can potentially reduce the amount of communication data and thus the delay, communication bandwidth and energy \cite{Duar12, Tan10, Tan10b}.
To realize such compression benefits,  the control center only needs to collect the reported data in $m_T < n_T$ RIs (i.e., compression over time) 
from a subset of $n_S$ nodes with size $m_S < n_S $ (i.e., compression over space) to reconstruct the complete data
for $n_S$ nodes in $n_T$ RIs. Note that data reconstruction can be performed locally at the data concentrator or at the control center. 
The later deployment is assumed in this work since it helps save bandwidth usage in the backhaul network.

\begin{figure*}[!t]
\centering
\includegraphics[width=125mm, height=50mm]{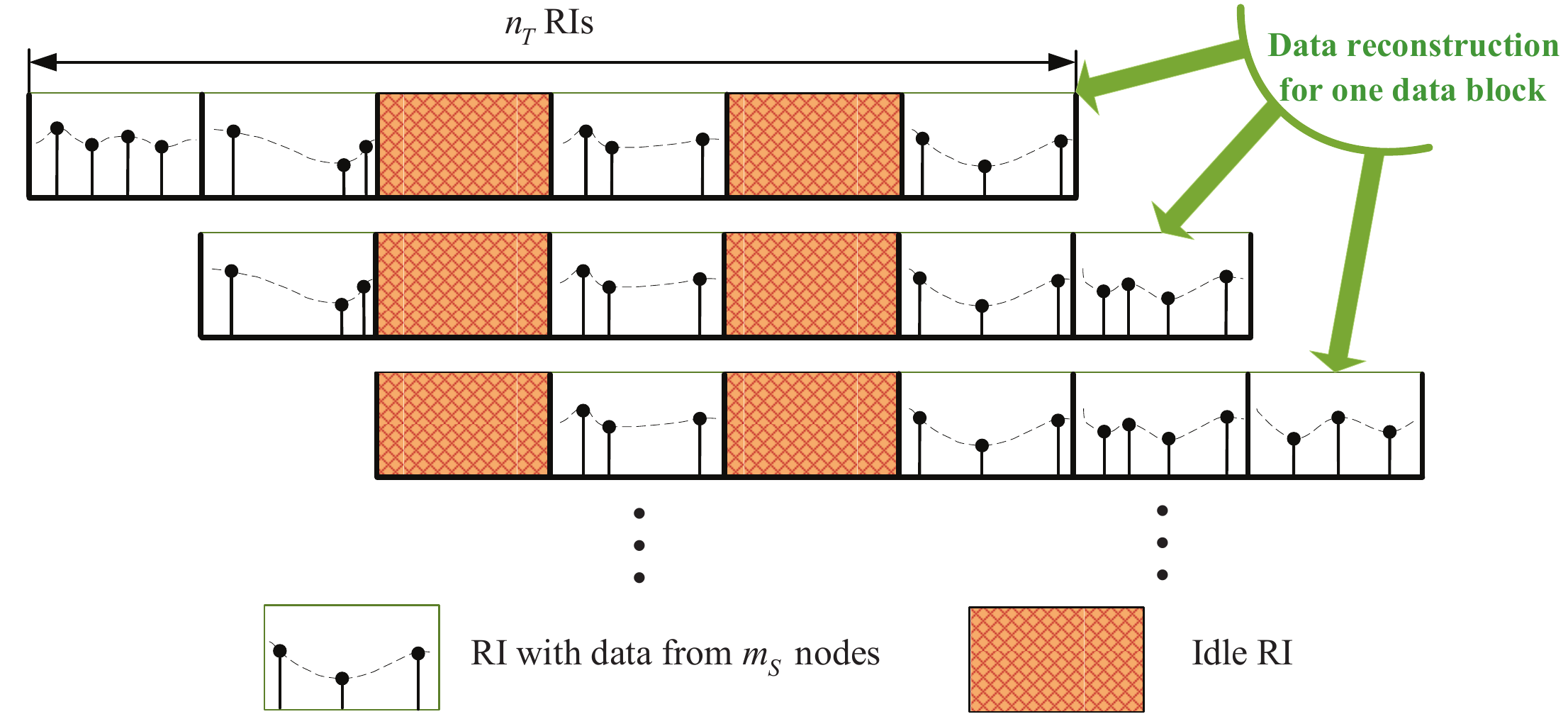}
\caption{Rolling-based data reconstruction for $n_S$ nodes over $n_T$ RIs}
\label{Rolling}
\end{figure*}

For practical implementation, the data transmission and reconstruction can be performed in a rolling
manner where data reconstruction performed at a particular RI utilizes the reported data over the latest $n_T$ RIs
as shown in Fig.~\ref{Rolling}. To guarantee the desirable data reconstruction quality, the control center must receive sufficient data which
is impacted by the underlying reporting mechanism and MAC protocol. Specifically, we must determine the values of 
$m_T$ and $m_S$ for some given values $n_T$ and $n_S$ to achieve the desirable data reconstruction reliability.
These design issues are addressed in the following sections.

\section{CS-Based Data Compression}
\label{GCSMAC}

\subsection{CS-Based Data Processing}

Without loss of generality, we consider data reconstruction of one data field for $n_T$ RIs and $n_S$ nodes.
Let $\mathbf{Z}$ be an $n_S \times n_T $ matrix whose $(i,j)$-th element denotes the injected
power at node $i$ and RI $j$. We will refer to the data from one RI (i.e., one column of $\mathbf{Z}$)
as one data block. From the CS theory, we can compress the data of interest if they possess 
sparsity properties in a certain domain such as wavelet domain. Specifically, we can express the
data matrix $\mathbf{Z}$ as
\beqn \label{wavtran}
\mathbf{Z} = \mathbf{\Psi}_S \mathbf{A} \mathbf{\Psi}_T^T
\eeqn
where  $\mathbf{\Psi}_S \in \mathbb{R}^{n_S \times n_S} $ and $\mathbf{\Psi}_T \in \mathbb{R}^{n_T \times n_T}$
denote wavelet bases in space and time dimensions, respectively \cite{Duar12, Tan10, Tan10b}.  The sparsity of $\mathbf{Z}$ can be
observed in the wavelet domain if matrix $\mathbf{A}$ has only $K$ significant (nonzero) coefficients where $K <n_S \times  n_T$.

We now proceed to describe the data compression and reconstruction operations. Let us denote
 $\mathbf{\Phi}_S \in \mathbb{R}^{m_S \times n_S} $ (for space) and $\mathbf{\Phi}_T \in \mathbb{R}^{m_T \times n_T}$ (for time) as the two 
sparse observation matrices where entries in these two matrices are i.i.d uniform random numbers where $m_S <  n_S $ and $m_T < n_T $.
Then we can employ  $\mathbf{\Phi}_S$  and $\mathbf{\Phi}_T$
to sample the power data from which we obtain the following observation matrix
\beqn
\mathbf{Y} = \mathbf{\Phi}_S \mathbf{Z} \mathbf{\Phi}_T^T.
\eeqn

Let $N_{\Sigma} = n_T n_S $ and $M = m_T m_S$ be the number of elements of $\textbf{Z}$ and $\mathbf{Y}$, respectively.
From the CS theory, we can reliably reconstruct the data matrix $\textbf{Z}$ by using the observation matrix $\mathbf{Y}$ if 
$m_T$ and $m_S$ are appropriately chosen. For the considered smartgrid communication design, this implies that
the control center only needs to collect $M$ injected power elements instead of $N_{\Sigma}$ values
for reliable reconstruction of the underlying data field. 

We now describe the data reconstruction for $\textbf{Z}$ by using the observation matrix $\mathbf{Y}$.
Toward this end, the control center can determine matrix $\mathbf{A}$, which corresponds
to the wavelet transform of the original data $\textbf{Z}$ as described in (\ref{wavtran}),
by solving the following optimization problem
\beqn
\min_ \textbf{A} \left\|\textbf{A}\right\|_2 \,\,\,
\text{s.t.}  \,\,\, \left\|\textbf{vec}\left(\mathbf{Y}\right) - \mathbf{\bar Y}\right\|_2 \leq \epsilon \label{OPT_RE_EQN}
\eeqn
where $\mathbf{\bar Y} = \left(\mathbf{\Phi}_S \otimes \mathbf{\Phi}_T \right) \left(\mathbf{\Psi}_S \otimes \mathbf{\Psi}_T \right) \textbf{vec}\left(\textbf{A}\right)$, $\otimes$ is the Kronecker product, $\textbf{vec} \left(\mathbf{X}\right)$ denotes the vectorization of the matrix $\mathbf{X}$, which stacks the rows of $\mathbf{X}$ into a single column vector. We can indeed solve problem (\ref{OPT_RE_EQN}) by using the Kronecker CS algorithm \cite{Duar12} to obtain $\mathbf{A}^* = \argmin_\textbf{A}  \left\|\textbf{A}\right\|_2$.
Then, we can obtain the estimation for the underlying data as $\textbf{vec}\left(\textbf{Z}^* \right) = \mathbf{\Psi}_S \otimes \mathbf{\Psi}_T \mathbf{A}^*$.
More detailed discussions of this data reconstruction algorithm can be found in \cite{Duar12}. 

Now there are two questions one must answer to complete the design: 1) how can one choose $m_S $ and $m_T$
to guarantee reliable data reconstruction for the underlying data field?; and 2) how can one design the data sampling
and MAC protocol so that the control center has sufficient information for data reconstruction? We will provide
the answers for these questions in the remaining of this paper.

\subsection{Determination of $m_S$ and $m_T$}
\label{GCSMAC11}

\begin{figure}[!t]
\centering
\includegraphics[width=80mm]{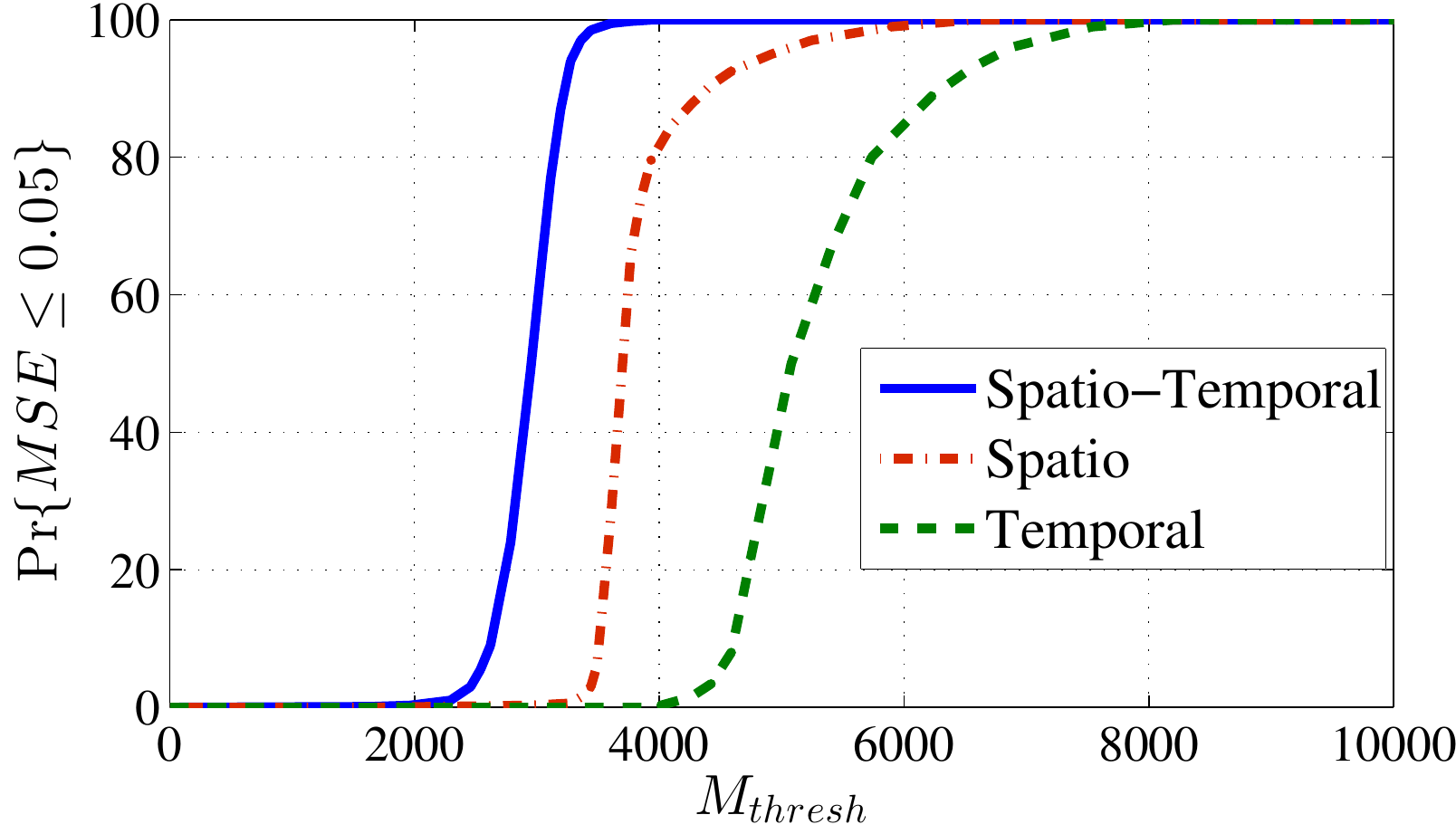}
\caption{Probability of success vs $M_{\sf thresh} $ with $n_S = 128 $ and $n_T = 128 $.}
\label{Novel_CDF_Numnode128_Numsample128}
\end{figure}

\begin{table}
\centering
\caption{Calculation of $M_{\sf thresh} $, $m_S $ and $m_T $}
\label{table}
\scriptsize
\setlength{\tabcolsep}{5pt}
\begin{tabular}{|c|c|c|c|c|c|c|}
\hline 
($n_S $, $n_T $) & (64,64) & (64,128) & (64,256) & (128,128) & (128,256) & (256,256)\tabularnewline
\hline
\hline 
$M_{\sf thresh} $ & 1551 & 1892 & 2880 & 3196 & 4620 & 9200 \tabularnewline
\hline 
($m_S $, $m_T $) & (33,47) & (22,86) & (16,180) & (47,68) & (30,154) & (80,115)\tabularnewline
\hline
\end{tabular}
\end{table} 

We would like to choose $m_S$ and $m_T$ so that $M = m_S m_T$
is minimum. Determination of the optimal values of $m_S$ and $m_T$ turns out to be a non-trivial task \cite{Duar12}.
So we propose a practical approach to determine $m_S$ and $m_T$. It is intuitive
that $m_S$ and $m_T$ should be chosen according to the compressibility in the space and time dimensions, respectively.
In addition, these parameters must be chosen so that the reliability of the data reconstruction meets the predetermined requirement,
which is quantified by the mean square error (MSE).

To quantify compressibility in the space and time, we consider two other design options, viz. temporal CS and spatial CS alone. 
For the former, the control center reconstructs data for one particular node by using the observations of only that node (i.e., we ignore spatial correlation).
For the later, the control center reconstructs the data in each RI for all nodes without exploiting the correlation over different RIs.
For fair quantification, we determine the MSE for one data field (i.e., for data matrix $\mathbf{Z}$ with $n_S \times n_T $ elements) for these two design options
where $MSE = \left\|\mathbf{Z}-\mathbf{Z}^*\right\|_2^2/\left\|\mathbf{Z}\right\|_2^2$.

For each spatial CS and temporal CS cases, we generate 1000 realizations of the injected powers based on which we perform 
the data reconstruction using the 1D CS for different
values of $m_S$ and $m_T$, respectively. Note that we randomly choose $m_S$ and $m_T$ out of $n_S$ nodes and $n_T$ RIs for spatial CS and temporal CS, respectively.
Then, we obtain the empirical probability of success for the data reconstruction
versus $M_{S} = m_S n_T$ and $M_{T} = n_S m_T$, respectively
where the ``success'' means that the MSE is less than the target MSE.
From the obtained empirical probability of success, we can find the required values of $M_S$ and $M_T$, which are denoted as $M_{S,{\sf thresh}}$ and $M_{T,{\sf thresh}}$, respectively, 
to achieve the target success probability.
Having obtained $M_{S,{\sf thresh}}$ and $M_{T,{\sf thresh}}$ capturing the compressibility in the
space and time as described above, we choose $m_S $ and $m_T$ for the 2D CS so that 
$m_S/m_T = n_S/n_T \times M_{S,{\sf thresh}}/M_{T,{\sf thresh}}$.
Similarly, we obtain the empirical probability of success for the 2D CS based on which 
we can determine the minimum value of $M_{\sf thresh} = m_S m_T$ to achieve the target success probability.

To obtain numerical results in this paper, we choose target $MSE = 0.05$ and target success probability equal 0.95.
Fig.~\ref{Novel_CDF_Numnode128_Numsample128} shows the empirical probability of success versus $M_{{\sf thresh}}= m_S m_T$ for 
$n_S = n_T = 128$ where  $M_{{\sf thresh}}$
in the horizontal axis represents the $M_{S,{\sf thresh}}$, $M_{T,{\sf thresh}}$, and $M_{{\sf thresh}}$ for 1D and 2D CS for simplicity.
The data model for the injected power will be described
in Section~\ref{Data_Model}. For this particular setting, we can obtain the ratio  $m_S/m_T = 0.691 $
from which can obtain the values of $m_S = 47 $ and $m_T = 68 $.

Similarly, we determine the $M_{\sf thresh}$, $m_S $ and $m_T$ for different scenarios whose corresponding $(n_S, n_T)$ values are given in Table~\ref{table}.
For all cases, we can observe that $m_S < n_S$ and $m_T < n_T$, which demonstrates the benefits of data compression by using the 2D CS. 
Having determined $m_S $ and $m_T$ as described above, the remaining tasks are to design
the DDR mechanism and MAC protocol that are presented in the following.

\section{DDR And MAC Protocol Design}
\label{Net_des_Per_Ana}

\subsection{Distributed Data Reporting}

In any RI, to perform reconstruction for the data field corresponding to the latest $n_T$ RIs, 
the control center must have data in $m_T$ RIs, each of which comprises $m_S$ injected powers
from $m_S$ nodes. With the previously-mentioned rolling implementation, the control center
can broadcast a message to all nodes to request one more data block (the last column of data matrix $\textbf{Z}$) if it has only $m_T-1$ data blocks 
to perform reconstruction in the current RI. At the beginning, the control center
can simply send $m_T$ broadcasts for $m_T$ randomly chosen RIs out of $n_T$ RIs and it performs data reconstruction
at RI $n_T$ upon receiving the requested data blocks.

\subsection{MAC Protocol Design}

\begin{figure*}[!t] 
\centering
\includegraphics[width=160mm, height=60mm]{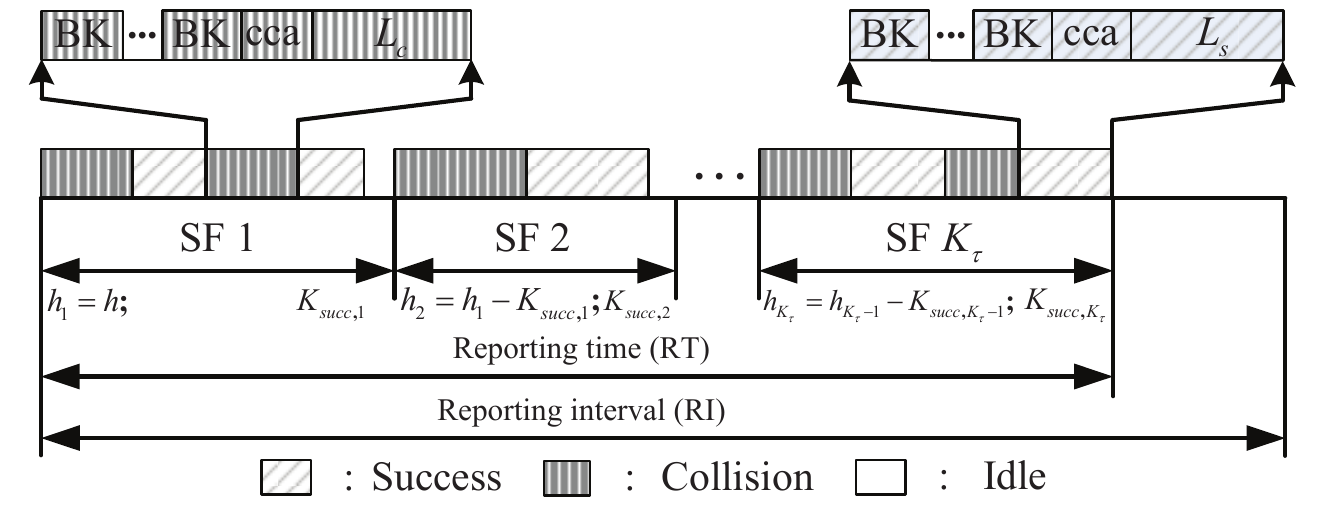}
\caption{Timing diagram of the MAC superframe structure for one RI.}
\label{cycletime}
\end{figure*}

If there is a broadcast message from the control center, we assume that each node participates in the contention process with probability $p_s$
using the slotted CSMA/CA MAC protocol in any RI. The slotted CSMA/CA MAC protocol \cite{Prime} with the proposed frame structure
is employed for data transmissions as described in the following. We set the optionally contention-free period to zero since we focus on distributed access design in this work.
Moreover, we assume there are $K_{\tau} $ superframes (SFs) in the contention period of any RI where $SF_i  = SF_0 \times 2^{BO_i}$ is the length of SF $i$,
 $SF_0 $ is the base length of SF, $BO_i \in \left[0,BO_{\sf max}\right]$ is the beacon order at SF $i$. Therefore, the reporting time (RT) in the underlying RI
is $ \sum_{i=1}^{K_{\tau}} SF_i$. We will optimize the parameters  $K_{\tau} $ and $\left\{BO_i\right\}$ to minimize the RT
while guaranteeing the desirable data reconstruction quality later. The superframe structure in one RI is illustrated in Fig.~\ref{cycletime}.

In each SF, the nodes that choose to access the channel (with probability $p_s$) perform the contention using the standardized slotted CSMA/CA protocol as follows.
A contending node randomly chooses a number in $\left[0, W_0\right]$ as the backoff counter ($W_0 \!=\! 2^{priority}$) and starts counting down.
If the counter reaches zero, the node will perform the clear channel  assessment (CCA) for $priority$ times (we set $priority \!=\! 2$). 
It will transmit data and wait for ACK if all CCAs are successful.
The reception of ACK is interpreted as a successful transmission, otherwise this is a collision.
In the case of failure in any CCA, the node attempts to perform backoff again with doubling backoff window.
In addition, each node is allowed to access channel up to $NB\!+\!1$ times. Since the length of the SF
is limited, a node may not have enough time to transmit its data and ACK packets at the end of the last SF.
In this case, we assume that the node will wait for the next SF to access the channel.
We refer to this as the deference state in the following.

\subsection{MAC Parameter Configuration for Delay Minimization}
\label{Net_design} 

We consider optimizing the MAC parameters $\left(K_{\tau}, p_s, \left\{BO_i\right\}\right)$ to minimize the reporting time in each RI
by solving the following problem:
\beqn
\label{OPT_DOST_EQN}
\begin{array}{l}
 {\mathop {\min }\limits_{K_{\tau}, p_s, BO_i}} \quad \mathcal{D}\left(K_{\tau}, p_s, \left\{BO_i\right\}\right) \\
 \mbox{s.t.} \quad \Pr\left\{K_{\text{succ}} \geq m_{S} \right\} \geq P_{\text{suff}},  \\
 \quad \quad  0 \leq K_{\tau} \leq K_{\tau,{\sf max}}, 0 \leq BO_i \leq BO_{\sf max}, 0 \leq p_s  \leq 1, \\
 \end{array}
\eeqn
where $\mathcal{D}(K_{\tau}, p_s, \left\{BO_i\right\}) = \sum_{i=1}^{K_{\tau}} SF_0 \times 2^{BO_i}$ is the reporting time.
Recall that $K_{\tau} $ is the number of SFs in one RI where $SF_i$ is the length of SF $i$ and $SF_0 $ is the base length of SF, $BO_i \in \left[0,BO_{\sf max}\right]$ is the beacon order at SF $i$, $BO_{\sf max}$ is the maximum value of $BO_i$, $p_s$ is the probability that each node performs a contention for a possible channel access.

In (\ref{OPT_DOST_EQN}), the first constraint means that the control center must receive $m_{S}$ packets (i.e., $m_{S}$ injected power values from $m_{S}$ nodes) 
with probability $P_{\text{suff}} \approx 1$. Note that one would not be able to deterministically guarantee the reception of $m_{S}$ packets due to the
random access nature of the DDR and MAC protocol.
The probability in this constraint can be written as
\beqn 
\Pr\left\{K_{\text{succ}} \geq m_{S} \right\} \!\! = \!\! \sum_{h= m_{S} }^{n_S} \!\! \Pr\left\{\widehat{m}_S = h\right\} \!\!\!\!\!\! \sum_{K_{\sf succ} = m_{S} }^h \sum_{l =1 }^{\left|\Xi\right|}  \prod_{i=1}^{K_{\tau}} \label{Prob_mthresh_OST_EQN_1}\\
\times \sum_{K_{S,i} = K_{{\sf succ},i}}^{K_{S,i,{\sf max}}} \!\!\!\!\!\!\!\! \Pr\!\left\{K_i = K_{S,i} \left|h_i \right.\right\} \! \Pr\!\left\{{\bar K}_i = K_{\text{succ},i} \left|K_{S,i},h_i \right.\right\} \label{Prob_mthresh_OST_EQN_3}
\eeqn
where $K_{\text{succ}}$ denotes the number of successfully transmitted packets in the RI and $\Pr\!\left\{\widehat{m}_S \!=\! h\right\} $ is the probability of $h$ nodes joining the contention.
Since each node decides to join contention with probability $p_s$, $\Pr\!\left\{\widehat{m}_S \!=\! h\right\} $ is expressed as 
\beqn
\label{Prob_htrans}
\Pr\left\{\widehat{m}_S = h\right\} = \left(\begin{array}{*{20}{c}} {n_S} \\  {h}  \\ \end{array}\right) p_s^h \left(1-p_s\right)^{n_S - h}.
\eeqn
In (\ref{Prob_mthresh_OST_EQN_1}) and (\ref{Prob_mthresh_OST_EQN_3}), we consider all possible scenarios so that the total number of
successfully transmitted packets over $K_{\tau}$ SFs is equal to  $K_{\text{succ}}$ where $K_{\text{succ}} \in \left[m_{S}, h\right]$. Here,
$K_{\text{succ},i}$ denotes the number of successfully transmitted packets in SF $i$ so that we have $\sum_{i=1}^{K_{\tau}} K_{\text{succ},i} = K_{\text{succ}}$.
In particular, we generate all possible combinations of $\left\{K_{\text{succ},i}\right\} $ for $K_{\tau} $ SFs and $\Xi $ represents the set of all possible combinations ($\left|\Xi\right|$ is the number of possible combinations). 

For each combination, we calculate the probability that the control center receives  $K_{\text{succ}}$ successful packets.
Note that  a generic frame may experience
one of the following events:  success, collision, CCA failure and deference.
Also, there are at most $K_{S,i,{\sf max}}$ frames in any SF $i$ where $K_{S,i,{\sf max}} = \left\lfloor SF_i/\min\left\{ NB+1, L_s+2 \right\}\right\rfloor $ since
the smallest length of a CCA failure frame is $NB +1 $ slots while the minimum length of a successful frame is $L_s+2$ where
$L_s$ is the required time for one successful transmission and 2 represents the two CCA slots.
We only consider the case that $K_{{\sf succ},i} \leq K_{S,i} \leq K_{S,i,{\sf max}}, \forall i \in \left[1, K_{\tau}\right]$.

In (\ref{Prob_mthresh_OST_EQN_3}), $\Pr\! \left\{K_i \!=\! K_{S,i} \left|h_i \right.\right\}$ is the probability that there are $K_{S,i}$ generic frames in SF $i$ given that $h_i$ nodes join
contention where $h_1 \!=\!h$ and $h_i \!=\! h_{i-1}\!-\! K_{\text{succ},i-1}$ since successfully transmitting node will not perform contention in the following frames.
Moreover, $\Pr\!\left\{{\bar K}_i \!=\! K_{\text{succ},i} \left|K_{S,i},h_i \right.\right\}$ is the probability that $K_{\text{succ},i}$ nodes transmit successfully in SF $i$ 
given that there are $h_i$ contending nodes and $K_{S,i}$ generic frames.

In order to calculate $\Pr\left\{K_i = K_{S,i} \left|h_i \right.\right\}$ and $\Pr\left\{{\bar K}_i = K_{\text{succ},i} \left|K_{S,i},h_i \right.\right\}$,
we have to analyze the Markov chain capturing detailed operations of the MAC protocol. 
For simplicity, we set $NB_i = NB = 5$, which is the default value. We analyze the Markov chain model in Appendix~ \ref{Mark_chain_mod}.
Then we determine $\Pr\left\{K_i = K_{S,i} \left|h_i \right.\right\} $ and $\Pr\left\{{\bar K}_i = K_{\text{succ},i} \left|K_{S,i},h_i \right.\right\} $ as follows.

\subsubsection{Calculation of $\Pr\left\{K_i = K_{S,i} \left|h_i \right.\right\} $}
\label{Pr_K_S_i1} 

In SF $i $ there are $h_i $ contending nodes and $K_{S,i}$ generic frames where frame $j$ has length $T_{ij}$.
We can approximate the distribution of generic frame length $T_{ij}$ as the normal distribution \cite{Park11}.
So the probability of having $K_{S,i} $ generic frames is written as
\beqn
\label{K_genfram_EQN}
\Pr\!\left\{\!K_i \!= \!K_{S,i} \left|h_i\right. \!\right\} \!\!=\!\! \Pr\{\sum_{j=1}^{K_{S,i}} \! T_{ij} \!\! = \!\! SF_i \} \!=\!\mathcal{Q} (\frac{SF_i\!-\!K_{S,i} {\bar T}_i}{\sqrt{K_{S,i} \sigma^2_i}})
\eeqn
where ${\bar T}_i $ and $\sigma^2_i $ are the average and variance of the generic frame length, respectively whose
calculations are presented in Appendix~\ref{average_variance}.

\subsubsection{Calculation of $\Pr\left\{{\bar K}_i = K_{\text{succ},i} \left|K_{S,i},h_i \right.\right\} $}
\label{Pr_K_S_i} 
The second quantity, $\mathcal{P} = \Pr\left\{{\bar K}_i = K_{\text{succ},i} \left|K_{S,i},h_i \right.\right\} $ is equal to
\beqn
\mathcal{P}  = \!\!\!\!\!\! \sum_{j = 0}^{K_{S,i}- K_{\text{succ},i}} \sum_{k= 0}^{1} \left(\!\!\!\begin{array}{*{20}{c}} {K_{S,i}} \\ { K_{\text{succ},i},j,k,l}  \\ \end{array} \!\!\!\right) \mathcal{P}_{\text{succ},h_i}^{ K_{\text{succ},i}}  \mathcal{P}_{\text{coll},h_i}^j \mathcal{P}_d^k \mathcal{P}_{\text{ccas},h_i}^{l} \label{P_2_OST_EQN}
\eeqn
where $K_{\text{succ},i}\!+\!j\!+\!k\!+\!l \!=\! K_{S,i}$; $j$, $k$, and $l$ represent the number of frames with collision, deference, and CCA failure, respectively. Moreover,
$\mathcal{P}_{\text{succ},h_i}$,  $\mathcal{P}_{\text{coll},h_i}$, $\mathcal{P}_{\text{ccas},h_i}$, and $\mathcal{P}_d$
denote the probabilities of success, collision, CCA failure, and deference, respectively, whose calculations are given in Appendix B.
In (\ref{P_2_OST_EQN}), we generate all possible combinations each of which has different numbers of success, collision, CCA failure, and deference frames. Also,
the product behind the double summation is the  probability of one specific combination.

\subsection{MAC Parameter Configuration Algorithm}

\begin{algorithm}[h]
\caption{\textsc{Optimization of MAC Parameters}}
\label{OPT_Delay}
\begin{algorithmic}[1]

\FOR {each value of $K_{\tau} \in [1,K_{\tau,{\sf max}}]$}

\FOR {each possible set $\left\{BO_i\right\}$}

\STATE Find optimal ${\bar p}_s $ as ${\bar p}_s = \mathop {\argmin} \limits_{0 \leq p_s \leq 1} \mathcal{D} \left( K_{\tau}, \left\{BO_i\right\}, p_s\right) $.

\ENDFOR

\STATE The best $\left(\left\{\bar{BO}_i\right\}, {\bar p}_s\right)$ for each $K_{\tau} $ is $\left(\left\{\bar{BO}_i\right\}, {\bar p}_s\right) = \mathop {\argmin} \limits_{\left\{{ BO}_i\right\}, {\bar p}_s} \mathcal{D} \left(K_{\tau}, \left\{BO_i\right\}, {\bar p}_s\right) $.
\ENDFOR

\STATE The final solution $\left( {\bar K}_{\tau}, \left\{\bar{BO}_i\right\}, {\bar p}_s  \right) $ is determined as $\left( {\bar K}_{\tau}, \left\{\bar{BO}_i\right\}, {\bar p}_s  \right) = \mathop {\argmin} \limits_{K_{\tau}, \left\{\bar{BO}_i\right\}, {\bar p}_s} \mathcal{D} \left(K_{\tau}, \left\{\bar{BO}_i\right\}, {\bar p}_s\right) $.

\end{algorithmic}
\end{algorithm}

The procedure for finding $\left(K_{\tau}, p_s, \left\{BO_i\right\}\right)$ can be described in Alg.~\ref{OPT_Delay}.
Since there are only finite number of possible choices for $K_{\tau} \in [1,K_{\tau,{\sf max}}]$ and the set $\left\{{BO}_i\right\}$,
we can search for the optimal value of $p_s$ for given $K_{\tau}$ and $\left\{{BO}_i\right\}$ as in step 3.
Then, we search over all possible choices of $K_{\tau}$ and the set $\left\{{BO}_i\right\}$ to determine the optimal
configuration of the MAC parameters (in steps 5 and 7).

\subsection{Bandwidth Usage}

To quantify the bandwidth usage, we consider a particular neighborhood with $N>n_S$ nodes, whose simultaneous transmissions
can collide with one another. In addition, these $N$ nodes must report injected power data to a control center. In this case, we would need
$N/n_S$ orthogonal channels\footnote{We ignore the fact that this number must be integer for simplicity} to support these communications. Suppose that 
the considered smartgrid application has a maximum target delay of $\mathcal{D}_{\sf max}$.
We should design the group size with $n_S$ nodes as large as possible while respecting this target delay in order
to minimize the required bandwidth (i.e., number of channels).

Let the maximum numbers of nodes for one group under TDMA and CSMA-CS schemes while still respecting the target delay be $n_S^{\sf TDMA}$ and $n_S^{\sf CSMA-CS}$, respectively.
Note that the TDMA scheme uses all $n_T$ RIs for data transmission while
 the CSMA-CS scheme only chooses $m_T$ RIs for each $n_T$ RIs to transmit the data.
Thus, the CSMA-CS scheme allows $n_T/m_T$ groups to share one channel for each interval of $n_T$ RIs.
As a result, the number of channels needed for $N$ nodes is $N/n_S^{\sf TDMA}$ for the TDMA scheme
and $N/n_S^{\sf CSMA-CS} \times m_T/n_T$ for the CSMA-CS scheme.

\subsection{Energy Consumption}

We now calculate the average energy consumption for data reporting of one data block. 
In each RI, there are $K_{\tau} $ SFs where the number of contending nodes decreases over the SFs.
Let $E_i$ denote the energy consumption per node in SF $i $ with $h_i $ contending nodes. The derivation of
$E_i$ is given in Appendix~ \ref{energy_consumption_i}.
Then, the average energy consumption in one RI can be expressed as follows:
\beqn
E = \sum_{h=1}^{n_S} \Pr\left\{\widehat{m}_S = h\right\} \sum_{K_{\sf succ} = m_{S}}^h \sum_{l =1 }^{\left|\Xi\right|} \prod_{i=1}^{K_{\tau}}  \sum_{K_{S,i} = K_{{\sf succ},i}}^{K_{S,i,{\sf max}}} \times  \hspace{0.55cm} \label{Energy_EQN_1}\\
\Pr\!\left\{K_i = K_{S,i} \left|h_i \right.\right\} \! \Pr\!\left\{{\bar K}_i = K_{\text{succ},i} \left|K_{S,i},h_i \right.\right\} \sum_{i=1}^{K_{\tau}} h_i E_i. \label{Energy_EQN_2}
\eeqn
Here, similar to derivation of $\Pr\left\{K_{\text{succ}} \geq m_{S}\right\}$ in Section~\ref{Net_design}, we generate all 
combinations of $\left\{h_i\right\}$, which represents the number of contending nodes in SF $i$.
For one such combination, we derive the total average energy in $K_{\tau}$ SFs. Note that
$\sum_{i=1}^{K_{\tau}} h_i E_i$ is the total average energy corresponding to $\left\{h_i\right\}$.
As a result, the average consumed energy for reporting one data field (with size $n_S \times n_T$) is $\mathcal{E} = m_T \times E$.

\vspace{10pt}
\section{Performance Evaluation and Discussion}
\label{Results}

\subsection{Data Modeling and Simulation Setting}
\label{Data_Model}

In order to evaluate the performance of our proposed data compression and MAC protocol, 
we synthetically generate the data for power loads and distributed generation powers by using available methods \cite{Hu11}--\cite{Chen10b}, \cite{Soares08}--\cite{Wei90}
because real-world data is not available. 
There are various probabilistic and stochastic methods to model power data in the literature. While independent and identically distributed (i.i.d.) normal
and Weibull \cite{Wei90} distributions are commonly used to generate these types of data, 
the considered power data generated once for every 5-minute RI in this work would be highly dependent. Therefore, 
autoregressive models \cite{Wei90}, which belong to the correlated time-series category,
 are selected to generate the required simulation data.

To model the load power, we employ the autoregressive moving average (ARMA) \cite{Wei90} as a time series model
which consist of two components, namely, deterministic and stochastic parts \cite{Soares08}--\cite{Christiaanse71}. Hence,
the load power can be expressed as
\beqn
\label{Timeseries_EQN}
X_t = X_t^d + X_t^s
\eeqn
where $t$ is the time index, $X$ commonly represents active load power ($P_l$) and reactive load power ($Q_l$),
 $X_t^d$ is the deterministic component of $X$ and $X_t^s$ is the stochastic part of $X$. 
The deterministic component, which captures the trend of data, is represented by the trigonometric functions \cite{Soares08}--\cite{Christiaanse71} as
\beqn
\label{deter_EQN}
X^d_t = \chi_0 + \sum_{i=1}^{m_h} \left(\chi_{\text{re},i} \text{sin}\left(\frac{2\pi k_i t}{288}\right)+ \chi_{\text{im},i}\text{cos}\left(\frac{2\pi k_i t}{288}\right)\right)
\eeqn
where $m_h$ is the number of harmonics (i.e., the number of trigonometric functions),
$\chi_{\text{re},i}$ and $\chi_{\text{im},i}$ are the coefficients of the harmonics, 
$\chi_0$ is the constant, and $k_i \leq 288/2$ for $\forall i \in \left\{1,\ldots, m_h\right\}$.
This form consists of a constant $\chi_0$ (i.e., the first quantity in (\ref{deter_EQN})) and $m_h$ trigonometric 
functions (i.e., the second quantity in (\ref{deter_EQN})). 
This general model indeed offers flexibility where we can define suitable formulas to present the effects of seasonality 
and special days which are extensively studied in \cite{Soares08, Papalexopoulos90}.
Now, the stochastic component is modeled by the first-order autoregressive process AR(1) which is the simple form of ARMA($n,m$) (i.e., $n = 1, m = 0$) \cite{Wei90}
as follows:
\beqn
\label{Stocha_EQN}
X_{t+1}^s = \varphi_t X_t^s + U_t
\eeqn
where $\varphi_t$ is the AR(1) coefficient and $U_t$ is the white noise process with zero mean
and variance of $(1-\varphi_t)$.
Here, we use AR(1) as an appropriate and simple data model; however other more complicated methods 
can be used as well \cite{Soares08, Papalexopoulos90, Wei90}.

We now present the method to determine the parameters ($\chi_0$, $\chi_{\text{re},i}$, $\chi_{\text{im},i}$, $m_h$, $\varphi_{t}$) in (\ref{deter_EQN}) and (\ref{Stocha_EQN}).
These parameters are coupled as indicated in (\ref{deter_EQN}) and (\ref{Stocha_EQN}). However we can separately estimate the parameters for
these deterministic and stochastic components without significantly increasing estimation errors compared to those due to joint estimation \cite{Soares08}.
To estimate these parameters, we use online data set \cite{UCIMLR} which represents the active and reactive load powers from 2006 to 2010.
Specifically, we use the ordinary least squares (OLS) \cite{Wei90} to estimate $\chi_0$, $\chi_{\text{re},i}$, and $\chi_{\text{im},i}$, 
$i = 1,\ldots, m_h$ \cite{Soares08, Christiaanse71}. Moreover, we employ the Bayesian information criterion (BIC) \cite{Wei90} to determine $m_h$ 
significant harmonics which is usually less than 5 for the selected data in \cite{UCIMLR}. 
Then we also use OLS algorithm to estimate the AR(1) coefficient, i.e., $\varphi_t$ for every 5-minute interval over one day \cite{Soares08, Christiaanse71}.
We should note that these parameters are updated and stored in every 5-minute interval over one day. 
Then we can generate the required power data in such the way that the output of the current interval is the input of the next interval.

We can similarly generate the distributed generation power data, which exhibits the temporal and spatial correlation.
In particular, distributed generation power data model also comprises deterministic and stochastic parts as described in (\ref{Timeseries_EQN}), (\ref{deter_EQN}) and (\ref{Stocha_EQN}). 
In addition, all parameters in these models are estimated by using online data of wind speeds \cite{KSCSW} which represent the 5-minute wind speeds in 2006--2010.
These data can be transformed to output power data as follows \cite{Chen10b, Papaefthymiou08}:
\beqn
S_g = \left\{ {\begin{array}{*{20}{c}}
   0 & {v \leq v_{ci}\,or\,v > v_{co}}  \\
   {A + Bv + Cv^2} & {v_{ci} < v \leq v_r}  \\
   {P_r} & {v_r < v \leq v_{co}}  
\end{array}} \right.
\eeqn
where $v_{ci}$ , $v_{co}$, $v_r$ and $P_r$ are the cut-in, cut-out, nominal wind speeds, and the rated power output, respectively.
These parameters and $(A, B, C)$ are determined as in \cite{Papaefthymiou05}.
Finally, the spatial correlation of distributed generation powers can be modeled as in \cite{Hu11}--\cite{Chen10b} 
where the correlation coefficient between 2 distributed generators $i$ and $j$ is $\rho_{i,j} = \exp\left(-d_{i,j}/d\right)$ 
where $d = 20 km$ is a positive constant and $d_{i,j}$ is the distance between generators $i$ and $j$, which is randomly chosen in $\left(0,1 km\right]$.

We assume that wind/solar generators are installed at a half of total nodes for all following experiments.
We ignore the node index $i$ in these notations for brevity.
The RI is set as $\tau_T = 5$ minutes. The target probability in the constraint (\ref{OPT_DOST_EQN}) is chosen as $P_{\text{suff}} = 0.9$.
The MAC parameters are chosen as  $L_s = T_p + t_{ACK} + L_{ACK}$, $T_p = 5+L_{MAC}$ slots ($L_{MAC} = 2$ is the MAC header), $L_{ACK} = 2$ slots, $t_{ACK} = 1$ slot, $t_{ACK,ti} = 4$ slots
where $T_p $ is the length of packet, $t_{ACK} $ is the idle time before the ACK, $L_{ACK} $ is the length of ACK, $t_{ACK,ti}$ is the timeout of the ACK, $BO_{\sf max} = 8$, $K_{\tau,{\sf max}} =10$. 
For all the results presented in this section, we choose $n_T=256$.

\subsection{Numerical Results and Discussion}

\begin{figure*}[!t]
\centering
\mbox{\subfigure[]{\includegraphics[width=2.20in]{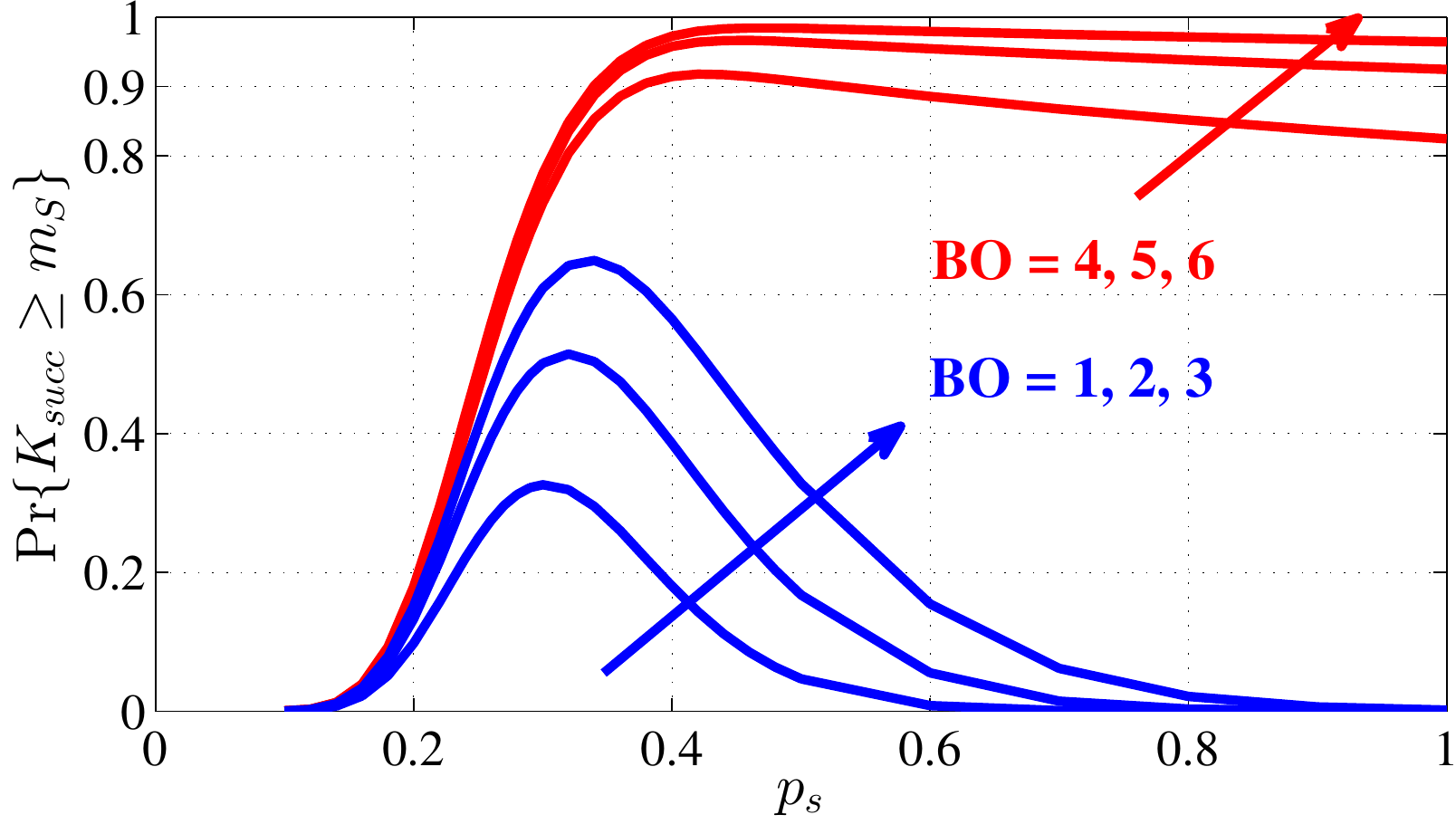} \label{Prsuff_Numnode_64_K_tau_3_BO_p_s1‎}}  
\subfigure[]{\includegraphics[width=2.20in]{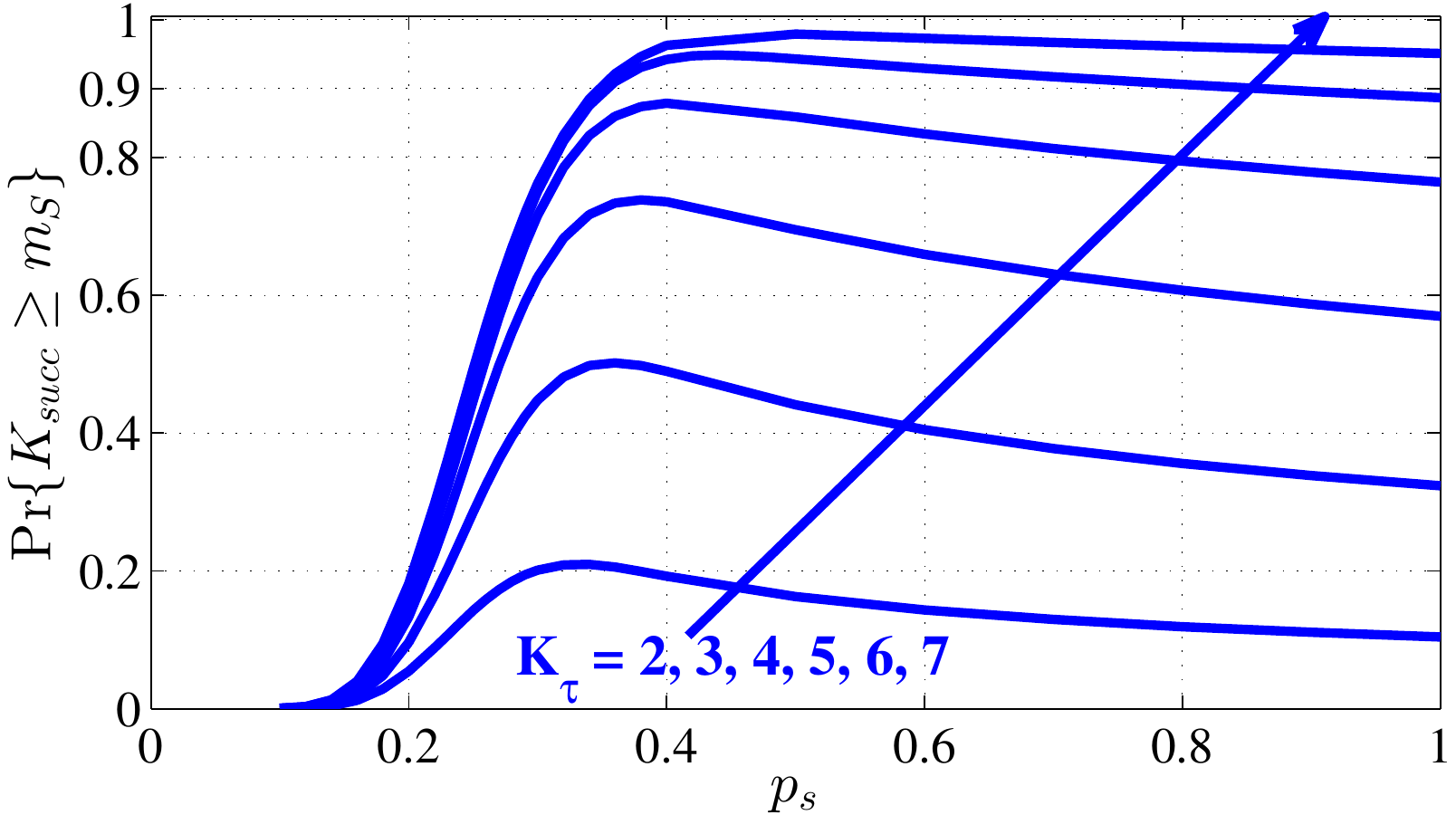} \label{Prsuff_Numnode_64_K_tau_BO_3_p_s2}} 
\subfigure[]{\includegraphics[width=2.20in]{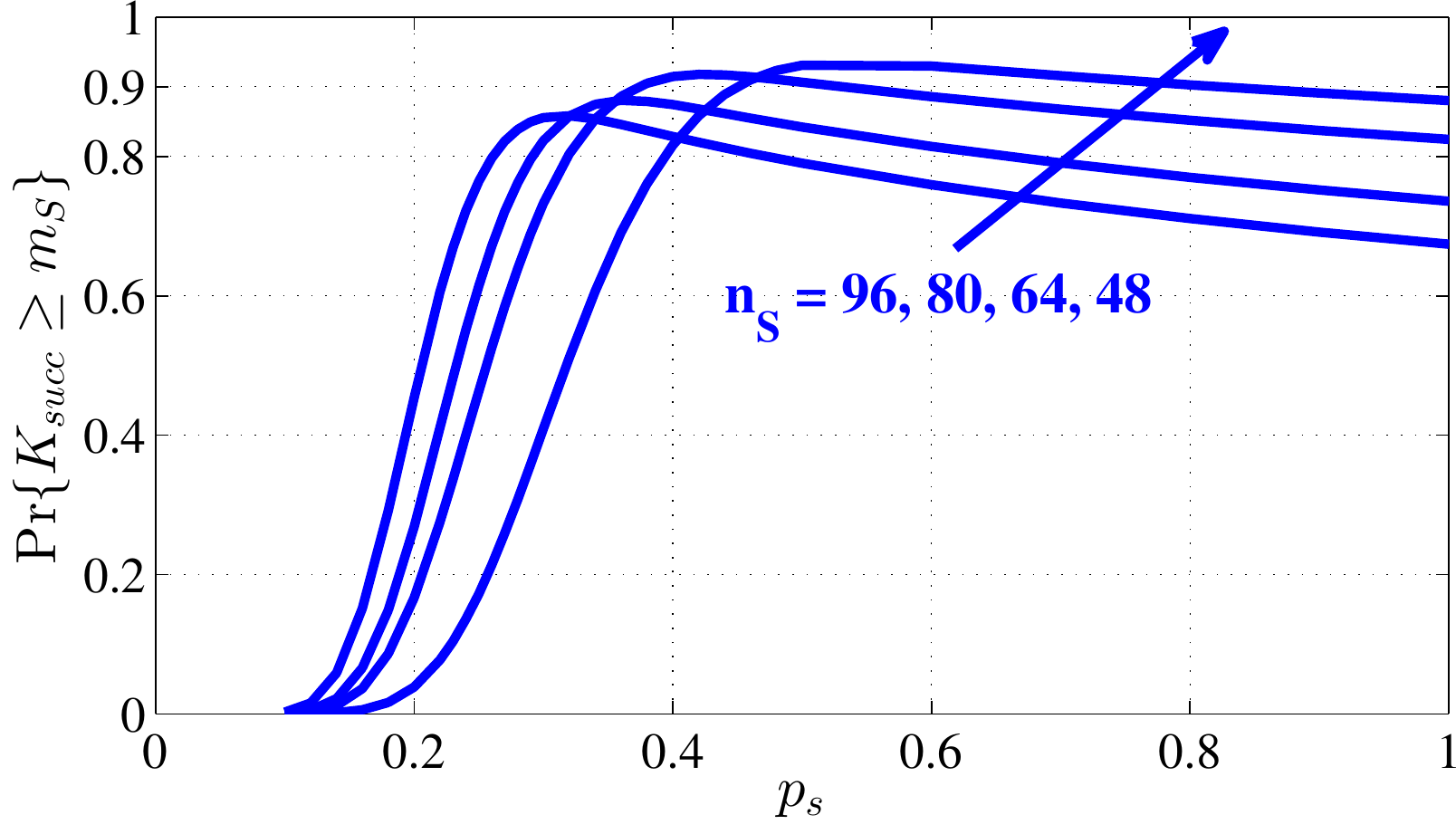} \label{Prsuff_Numnode_K_tau_3_BO_4_p_s}} }
\caption{$\Pr\left\{K_{\text{succ}} \geq m_{S}\right\} $ vs. $p_s$ for (a) $(n_S, m_{S}, K_{\tau}) = (64, 16, 3)$ and
 different values of $BO$,  (b) $(n_S, m_{S}, BO) = (64, 16, 3)$ and different values of $K_{\tau}$, (c) $(BO, K_{\tau}) = (4, 3) $ and different values of $n_S $.}
\end{figure*}

\subsubsection{Sufficient Probability}

In Fig.~\ref{Prsuff_Numnode_64_K_tau_3_BO_p_s1‎}, we show the variations of sufficient probability, namely $\Pr\!\left\{K_{\text{succ}} \!\geq\! m_{S}\right\}$, versus 
$p_s$ for different values of $BO_i \!=\! BO $ (i.e., all SFs employ the same $BO$) where $m_{S} \!=\! 16$, $K_{\tau} \!=\! 3$,
 and $n_S \!=\! 64$. It can be observed that there exists an optimal $p_s$ that maximizes $\Pr\!\left\{K_{\text{succ}} \!\geq\! m_{S}\right\}$ for each value of $BO$.
This optimal value is in the range $\left[0.3, 0.5\right]$. Furthermore, 
$BO$ must be sufficiently large ($BO \!\geq\! 4$) to meet the required target value $P_{\text{suff}}$=0.9. In addition, larger values of $BO$ lead
 to longer SF length, which implies that more data packets can be transmitted.
In Fig.~\ref{Prsuff_Numnode_64_K_tau_BO_3_p_s2}, we show the probability $\Pr\left\{K_{\text{succ}} \geq m_{S}\right\} $ versus $p_s $ for
different values of $K_{\tau} $ where we set $n_S = 64 $ and $BO = 3$. This figure confirms that the maximum $\Pr\left\{K_{\text{succ}} \geq m_{S}\right\} $
becomes larger with increasing $K_{\tau} $ where we can meet the target probability $P_{\text{suff}}$=0.9 if $K_{\tau} \geq 6$.

Now we  fix $BO_i = BO = 4 $, $K_{\tau} = 3 $, then we study $\Pr\left\{K_{\text{succ}} \geq m_{S}\right\}$ versus $p_s $ for different values of $n_S $. 
Note that $m_{S}$ can be calculated for each value of $n_S $ as presented in Section~\ref{GCSMAC11}. Specifically, $m_{S}$ is equal to 13, 16, 19, and 22 for $n_S = 48, 64, 80 $, and 96, respectively.
Fig.~\ref{Prsuff_Numnode_K_tau_3_BO_4_p_s} demonstrates that smaller number of nodes $n_S$ (e.g., $n_S = 48, 64$) can achieve the target 
value of $P_{\text{suff}} $ at the optimal $p_s$.
However, as $n_S$ increases to 80 and 96, there is no $p_s $ that meets the target value of $P_{\text{suff}}$.
This is because the collision probability is lower for the smaller number of nodes, which results in larger values of $\Pr\left\{K_{\text{succ}} \geq m_{S}\right\}$. 

\subsubsection{Reporting Delay}

\begin{figure}[!t] 
\centering
\includegraphics[width=80mm]{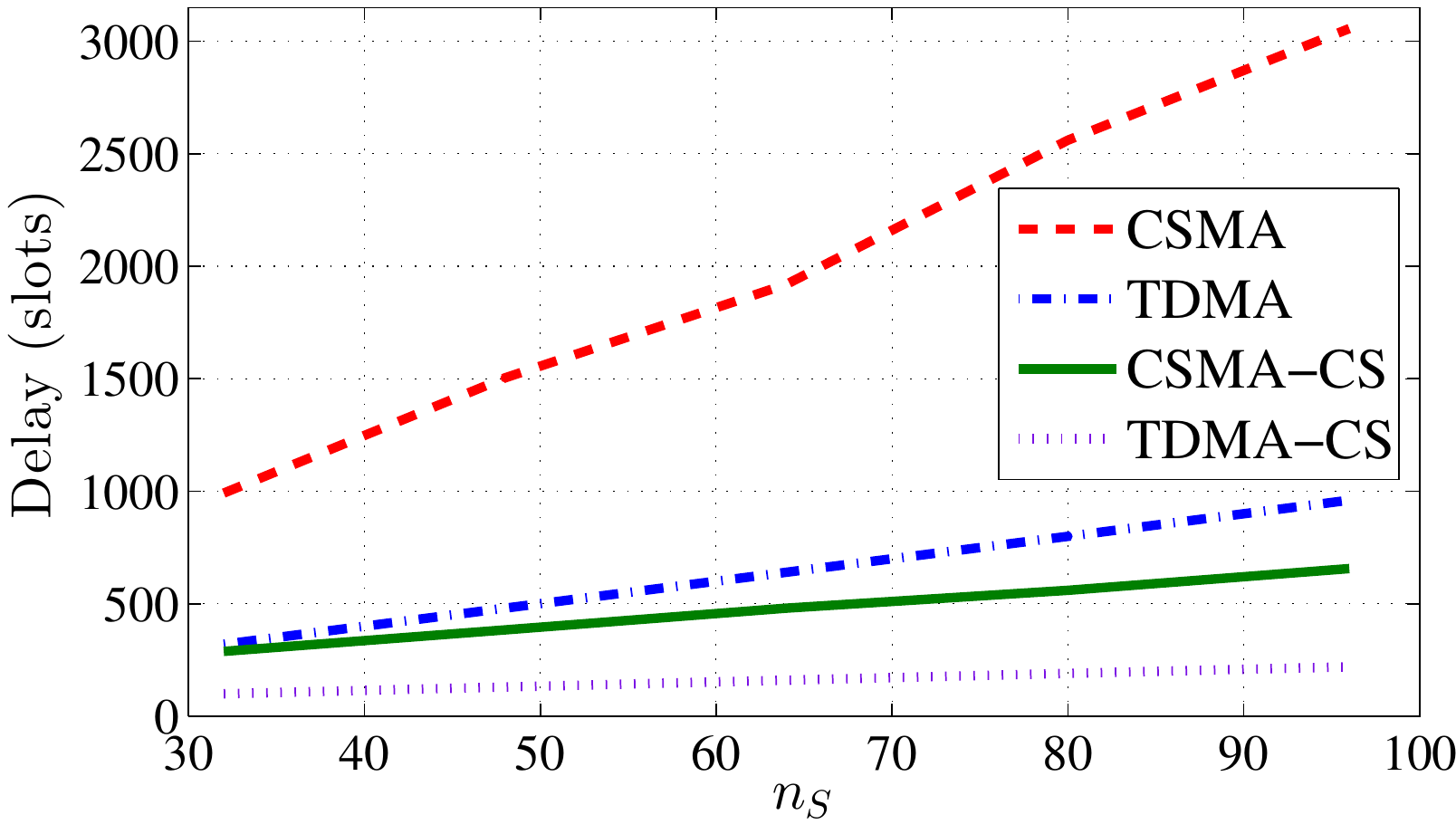}
\caption{Reporting delay $\mathcal{D}$ vs. $n_S$.}
\label{Delay_ns}
\end{figure}

We now show the optimal reporting delay $\mathcal{D}$ in one RI versus the number of nodes $n_S$ for different schemes,
namely TDMA, CSMA, TDMA-CS, and our proposed CSMA-CS schemes in Fig.~\ref{Delay_ns}. Here, X-CS refers 
to scheme X that integrates the CS-based data compression and X refers to scheme X without data compression.
Moreover, TDMA is the centralized non-contention time-division-multiple-access MAC, which always achieves better performance that the CSMA scheme.
For both CSMA and CSMA-CS schemes, their MAC parameters are optimized by using Alg.~\ref{OPT_Delay}.
It can be seen that our proposed CSMA-CS protocol achieves much smaller delay than the CSMA scheme, which
confirms the great benefits of employing the CS. In addition, TDMA-CS outperforms
our CSMA-CS protocol since TDMA is a centralized MAC while CSMA is a randomized distributed MAC.
Finally, this figure shows that our CSMA-CS protocol achieves better delay performance than 
the TDMA scheme. 

\begin{figure}[!t] 
\centering
\includegraphics[width=80mm]{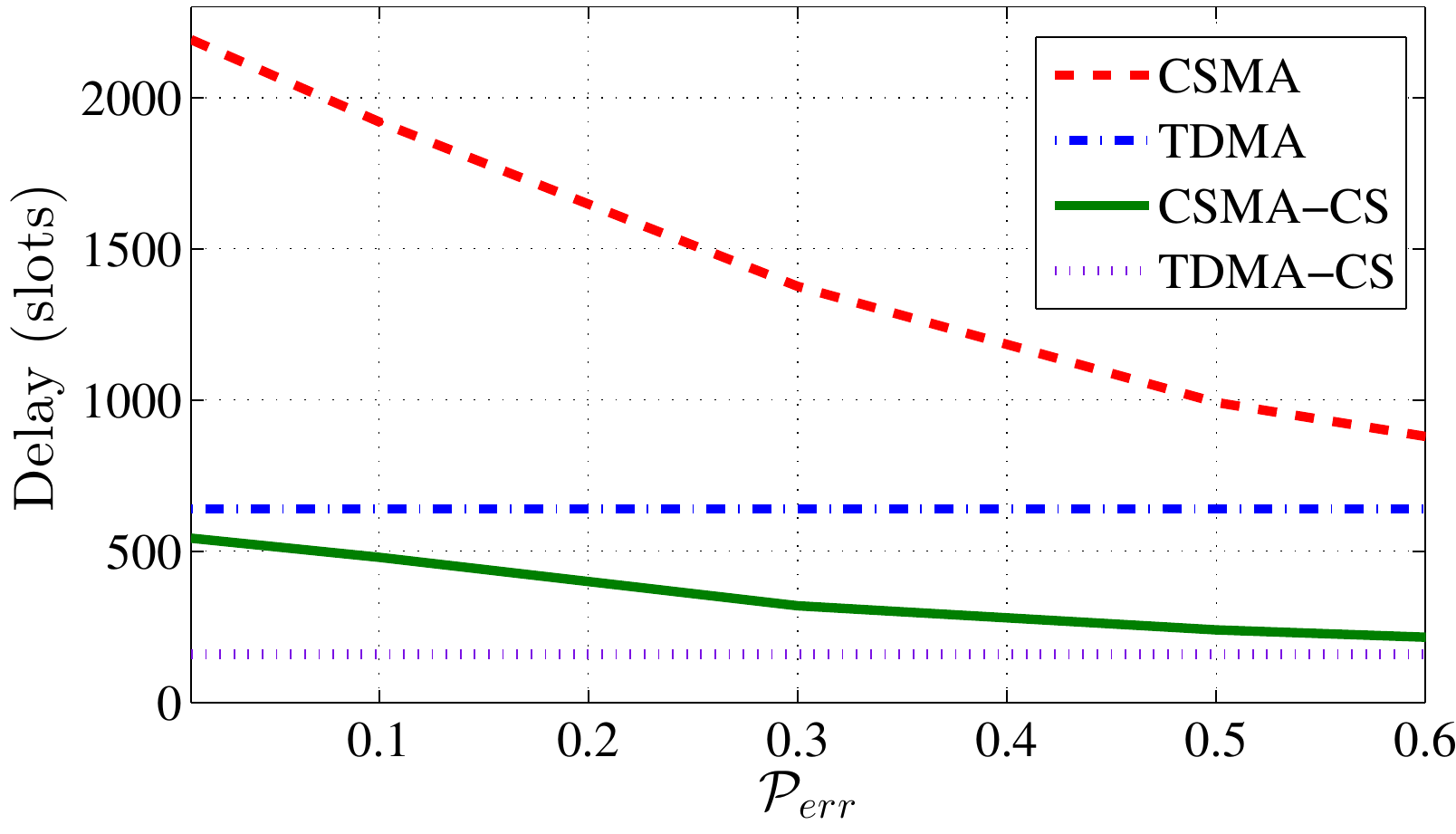}
\caption{Reporting delay $\mathcal{D}$ vs. $\mathcal{P}_{err}$.}
\label{Delay_P_err_n_S_64}
\end{figure}

We illustrate the variations of the optimal reporting delay with $\mathcal{P}_{err} $ for different schemes where 
$\mathcal{P}_{err} \!=\! 1\!-\!\Pr\!\left\{K_{\text{succ}} \!\geq\! m_{\text{S}}\right\}$
and $n_S \!=\! 64$ in Fig.~\ref{Delay_P_err_n_S_64}. This figure shows that as $\mathcal{P}_{err} $ increases, the reporting delay decreases.
This indeed presents the tradeoff between the reporting delay and $\mathcal{P}_{err}$. 
Note that the delay of the TDMA-CS scheme is the lower bound for all other schemes. Interestingly, as $\mathcal{P}_{err}$ increases
the delay gap between the proposed CSMA-CS and the TDMA-CS schemes becomes smaller. 

\begin{figure}[!t] 
\centering
\includegraphics[width=80mm]{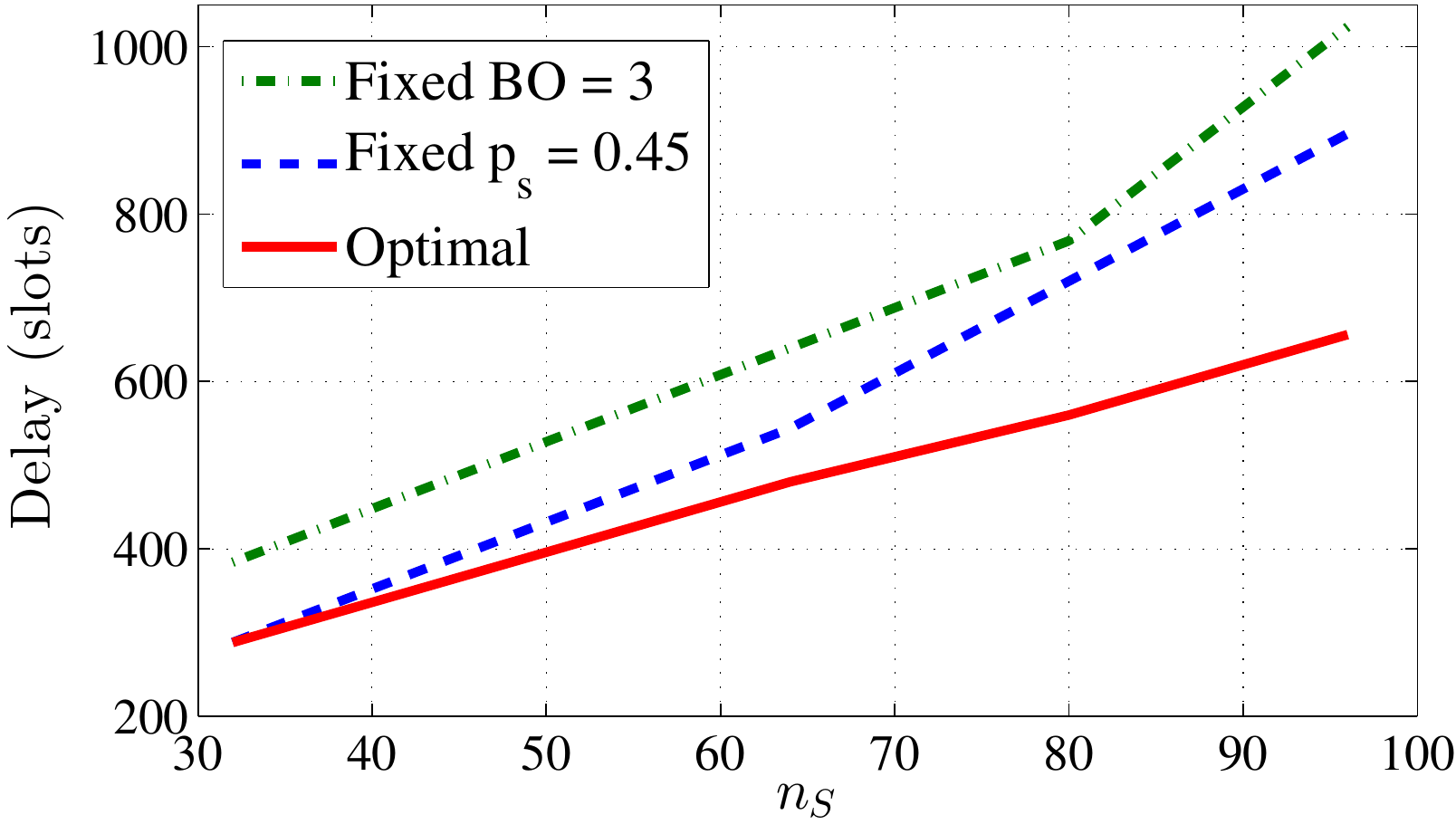}
\caption{Optimal $\mathcal{D}$ vs.  $n_S$ with full and partial optimizations.}
\label{minDelay_Numnode_BO_3p_s0_45}
\end{figure}

We compare the delay performance with partial and full optimization of the MAC protocol for our proposed scheme.
Specifically, we consider two following cases in which we set default value for one of three optimization parameters
 i) $BO_i = BO = 3$ for all SFs; ii) $p_s = 0.45 $.
For these two cases, we optimize the remaining parameters to achieve minimum reporting delay in each RI.
Fig.~ \ref{minDelay_Numnode_BO_3p_s0_45} shows that our proposed CSMA-CS protocol with full optimization outperforms the others
with the partial optimization. Furthermore, the performance for case i) with fixed $BO $ is pretty poor since 
 we only choose the optimal configuration for $p_s $ and $K_{\tau} $ to minimize the delay performance in this case.
However, the delay performance degrades more moderately as we fix the access parameter at $p_s = 0.45$.

\begin{figure}[!t] 
\centering
\includegraphics[width=80mm]{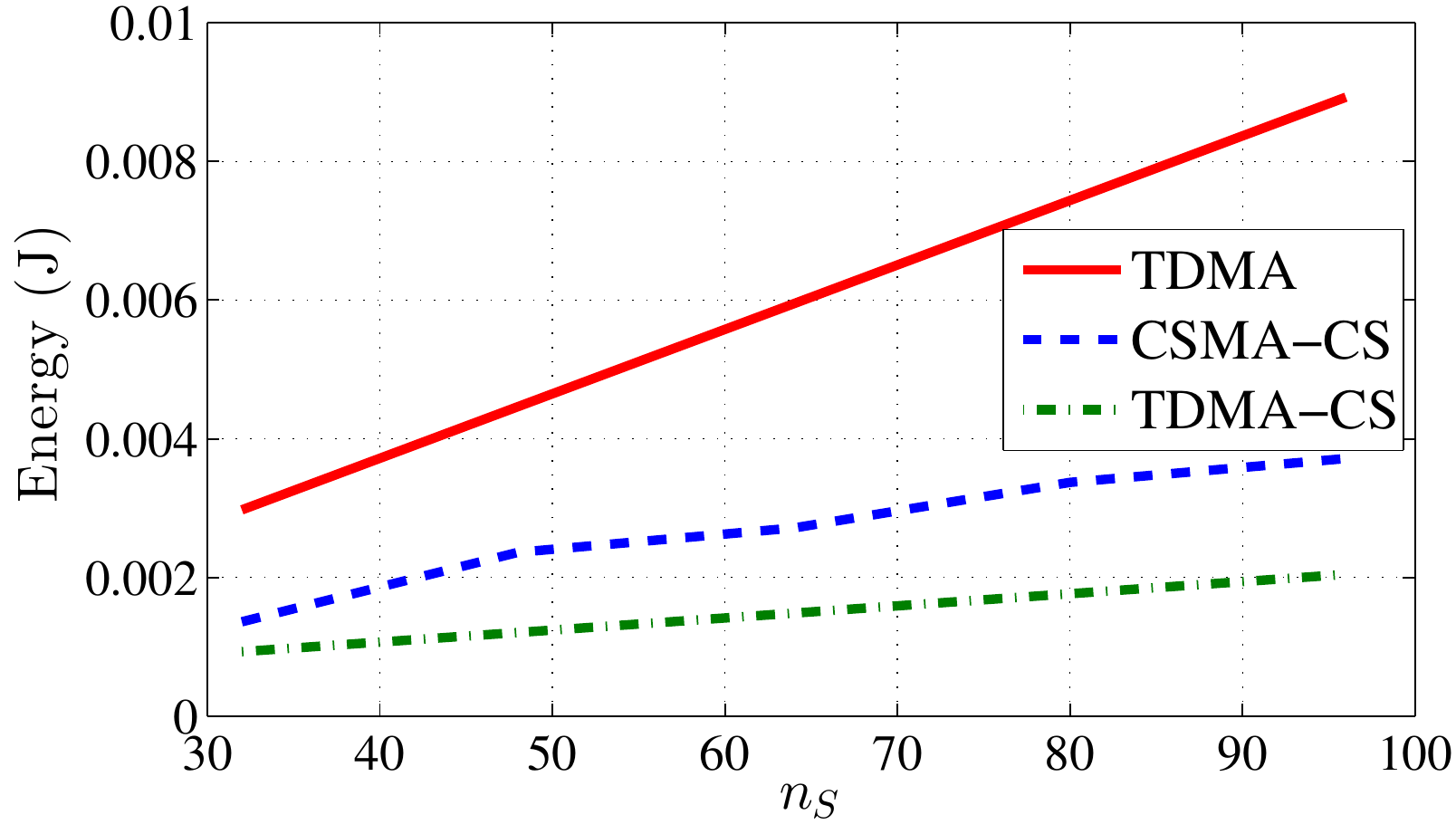}
\caption{Energy consumption $E$ vs. $n_S$ for one RI.}
\label{minEnergy_Numnode_K_tauBOp_s}
\end{figure}

\begin{figure}[!t] 
\centering
\includegraphics[width=80mm]{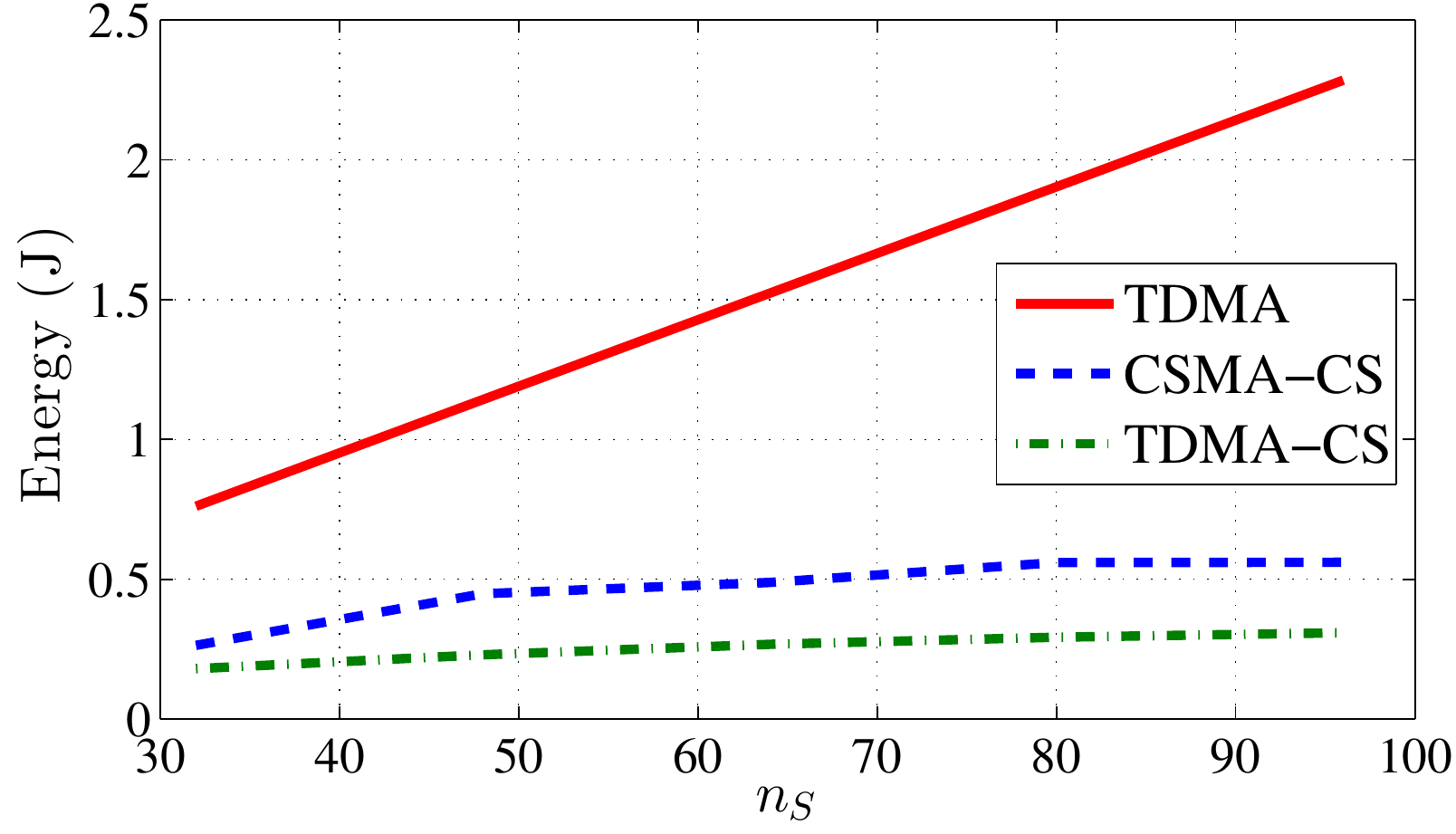}
\caption{Energy consumption $\mathcal{E}$ vs. $n_S$ for one data field.}
\label{minEnergy_onefield_Numnode_K_tauBOp_s}
\end{figure}

\subsubsection{Energy Consumption}

We present the energy consumption for different schemes in Figs.~\ref{minEnergy_Numnode_K_tauBOp_s}, \ref{minEnergy_onefield_Numnode_K_tauBOp_s}
under the following parameter setting: $E_{\sf idle} = 0.228 \mu J/{\sf slot}$, $E_{\sf tx} = 10.022 \mu J/{\sf slot}$, $E_{\sf rx} = 11.290 \mu J/{\sf slot}$, 
and $E_{\sf sens} =E_{\sf rx}$ \cite{Poll08}.
Fig.~\ref{minEnergy_Numnode_K_tauBOp_s} demonstrates the energy consumption $E$ in one RI for different values of $n_S$.
It can be seen that the proposed CSMA-CS scheme results in significant energy saving compared to the TDMA scheme.
For completeness, we also show the energy consumption for one data field (i.e., for $n_T$ RIs and $n_S$ nodes)
 in Fig.~\ref{minEnergy_onefield_Numnode_K_tauBOp_s}. Recall that in the proposed framework,
only data blocks corresponding to  $(m_S, m_T)$ are reported where $(m_S, m_T)$ can be determined for each $(n_S, n_T)$ as described in Section~ \ref{GCSMAC11}. 
Specifically, we have $(m_S, m_T) = \left\{(10, 194), (13, 190), (16, 180), (19, 166), (22, 151)\right\}$ for $n_T$=256 and 
$n_S = \left\{32, 48, 64, 80, 96\right\}$, respectively. Again, this figure confirms that the proposed CSMA-CS scheme outperforms the TDMA scheme.
Moreover, the consumed energy of the proposed CSMA-CS scheme is slightly higher than that due to the TDMA-CS scheme.

\begin{figure}[!t] 
\centering
\includegraphics[width=80mm]{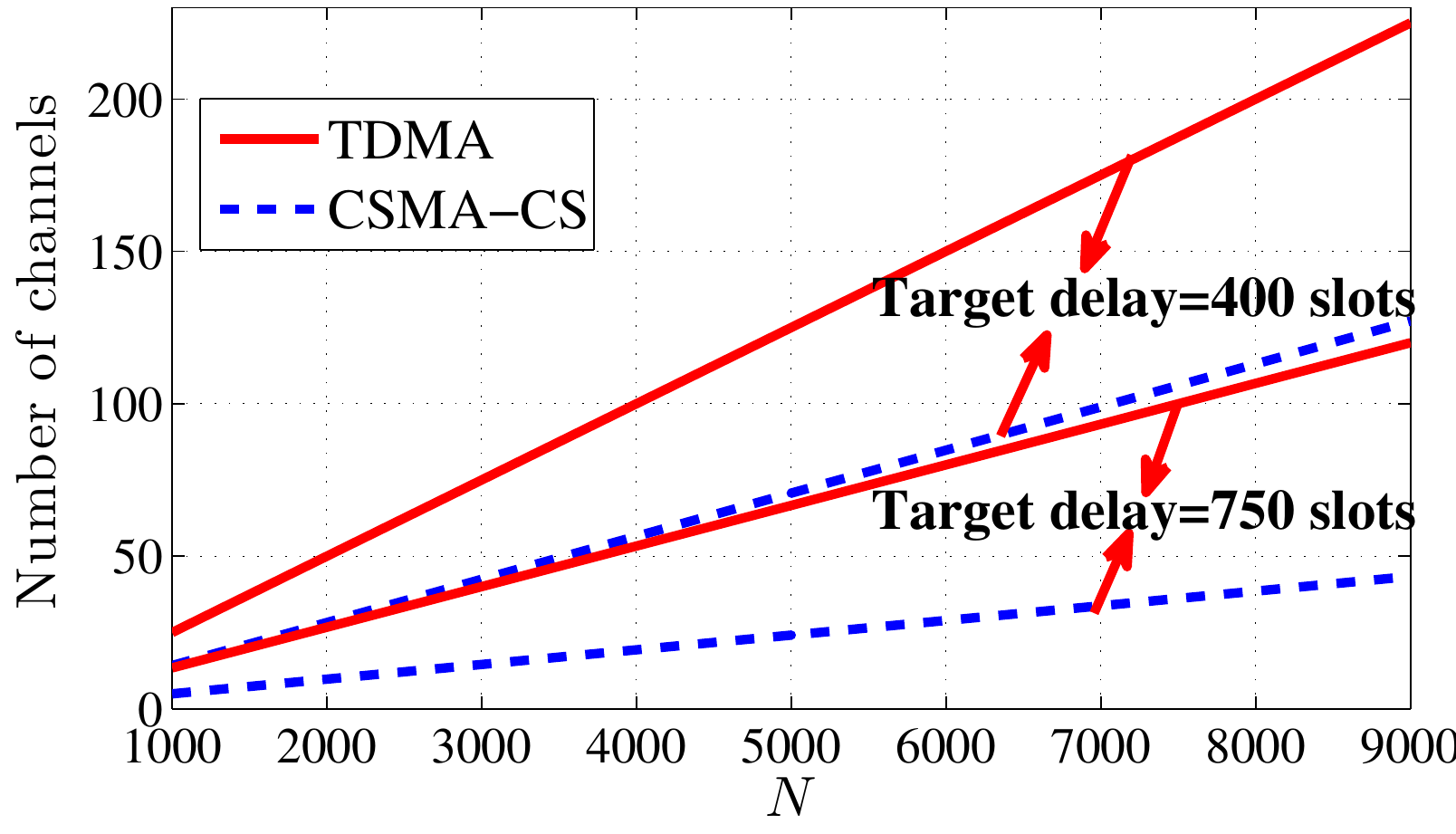}
\caption{Number of channels vs. $N$.}
\label{Numchan_N_Numnode_BOp_s}
\end{figure}

\begin{figure}[!t] 
\centering
\includegraphics[width=80mm]{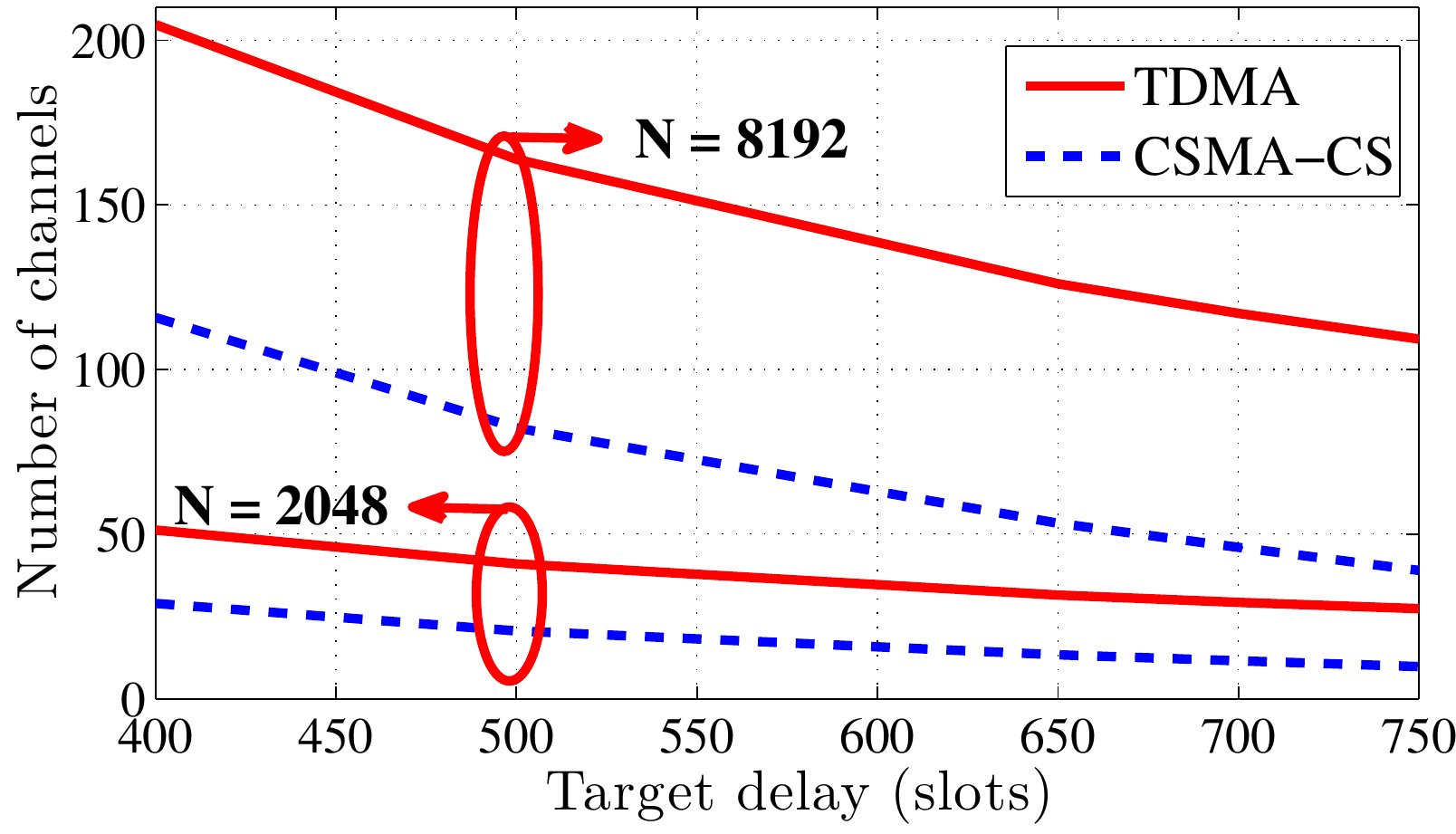}
\caption{Number of channels vs. target delay.}
\label{Delay_target_Numnode}
\end{figure}

\begin{figure}[!t] 
\centering
\includegraphics[width=80mm]{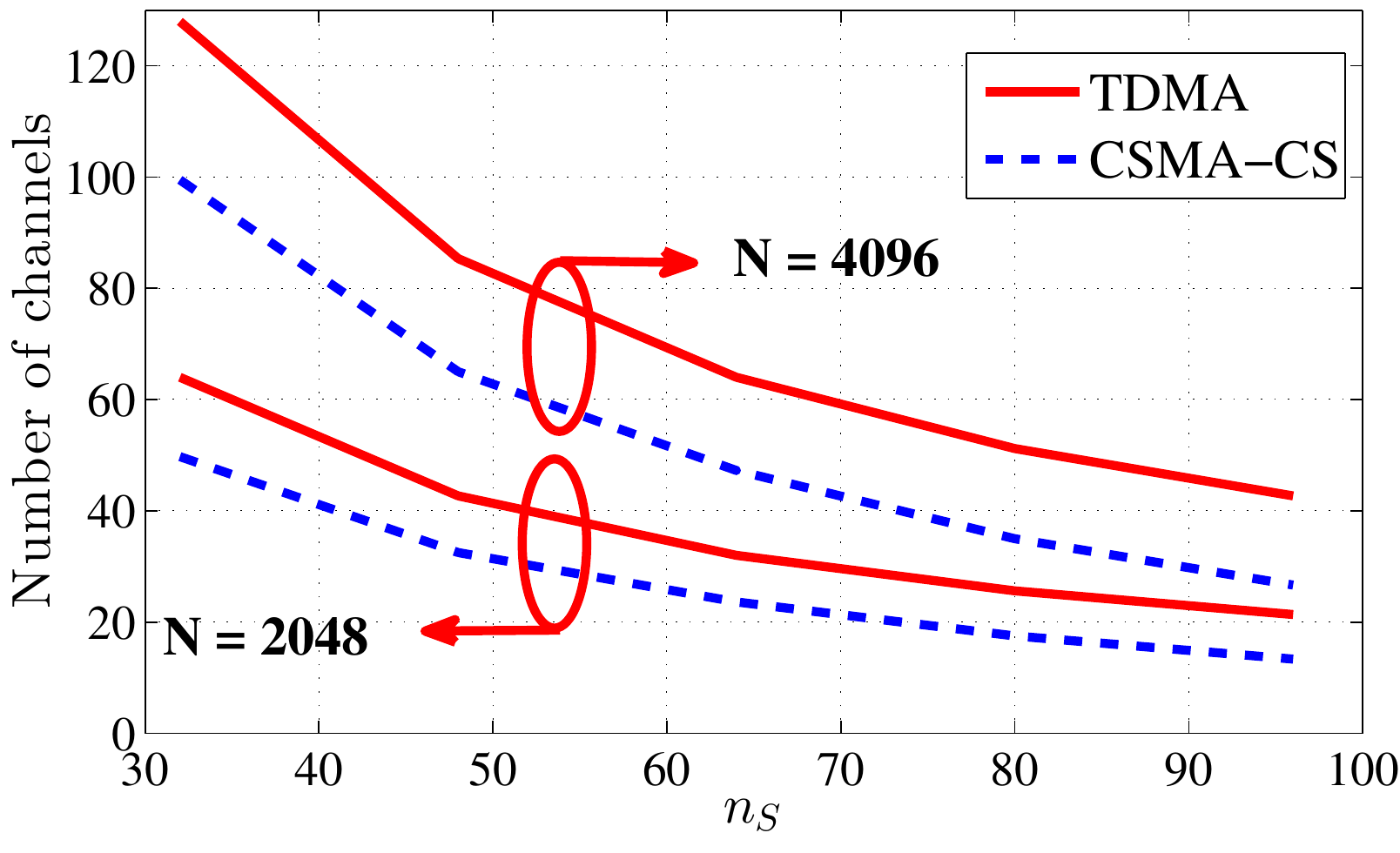}
\caption{Number of channels vs. $n_S$.}
\label{Numchan_nS_Numnode_BOp_s}
\end{figure}

\subsubsection{Bandwidth Usage}

We now examine the bandwidth usage due to different schemes. To ease the exposition, we do not show the 
results of the CSMA and TDMA-CS schemes. 
The group sizes for TDMA and CSMA-CS protocols, namely $n_S^{\sf TDMA}$, $n_S^{\sf CSMA-CS}$,
can be determined for a given target delay $\mathcal{D}_{\sf max}$ by using the results shown in Fig.~\ref{Delay_ns}.
Specifically,  we have $n_S^{\sf TDMA}\!\! = \left\{40, 75\right\}$ and $n_S^{\sf CSMA-CS}\!\! = \left\{55, 128\right\}$ for TDMA and CSMA-CS schemes
for the target delay values of $\mathcal{D}_{\sf max} = \left\{400, 750\right\}$ slots, respectively.
Then we calculate the required bandwidth (i.e., number of channels) for TDMA and CSMA-CS schemes with a given number of nodes $N$.

In Fig.~\ref{Numchan_N_Numnode_BOp_s}, we show the required number of channels versus $N$.
It can be observed that our proposed CSMA-CS scheme requires less than half of the bandwidth demanded by the TDMA scheme.
Also when the network requires smaller target delay, we need more channels for both schemes as expected.
Finally, Fig.~\ref{Delay_target_Numnode} illustrates the variations in the number of channels
 versus the target delay for $N = \left\{2048, 8192\right\}$.
Again, our proposed CSMA-CS scheme provides excellent bandwidth saving compared to the TDMA scheme.

In Fig.~\ref{Numchan_nS_Numnode_BOp_s}, we show the required number of channels versus $n_S$ for $N = \left\{2048, 4096\right\}$.
These results can be obtained as follows. Suppose that the target delay in one RI is 650 slots  then  the group sizes for TDMA and CSMA-CS protocols 
can be determined from the results in Fig.~\ref{Delay_ns} as $\left\{n_S^{\sf TDMA} , n_S^{\sf CSMA-CS}\right\} = \left\{65, 96\right\}$.
Then using the results in Fig.~\ref{Numchan_nS_Numnode_BOp_s}, the numbers of required channels for TDMA and CSMA-CS protocols are $\left\{64, 25\right\}$ for $N = 4096$ and
 $\left\{32, 13\right\}$ for $N = 2048 $. We can observe that the number of required channels for our proposed scheme is always less than that due to the TDMA protocol,
which again demonstrates the efficacy of our proposed design.

\vspace{0.2cm}
\section{Conclusion}
\label{Conclusion} 

We have proposed the joint design of data compression using the CS technique and CSMA MAC protocol for smartgrids
with renewable energy. We have shown how to choose the compression levels in the space and time dimensions to maintain desirable
reconstruction performance. Then, we have presented the design and optimization of the MAC protocol to minimize the reporting
delay. Furthermore, we have derived the bandwidth usage and energy consumption for our proposed scheme.
Numerical results have confirmed the significant performance gains of the proposed design compared to other non-compressed solutions.

\appendices

\section{Markov Chain Model for CSMA Protocol}
\label{Mark_chain_mod} 

\begin{figure*}[!t] 
\centering
\includegraphics[width=120mm]{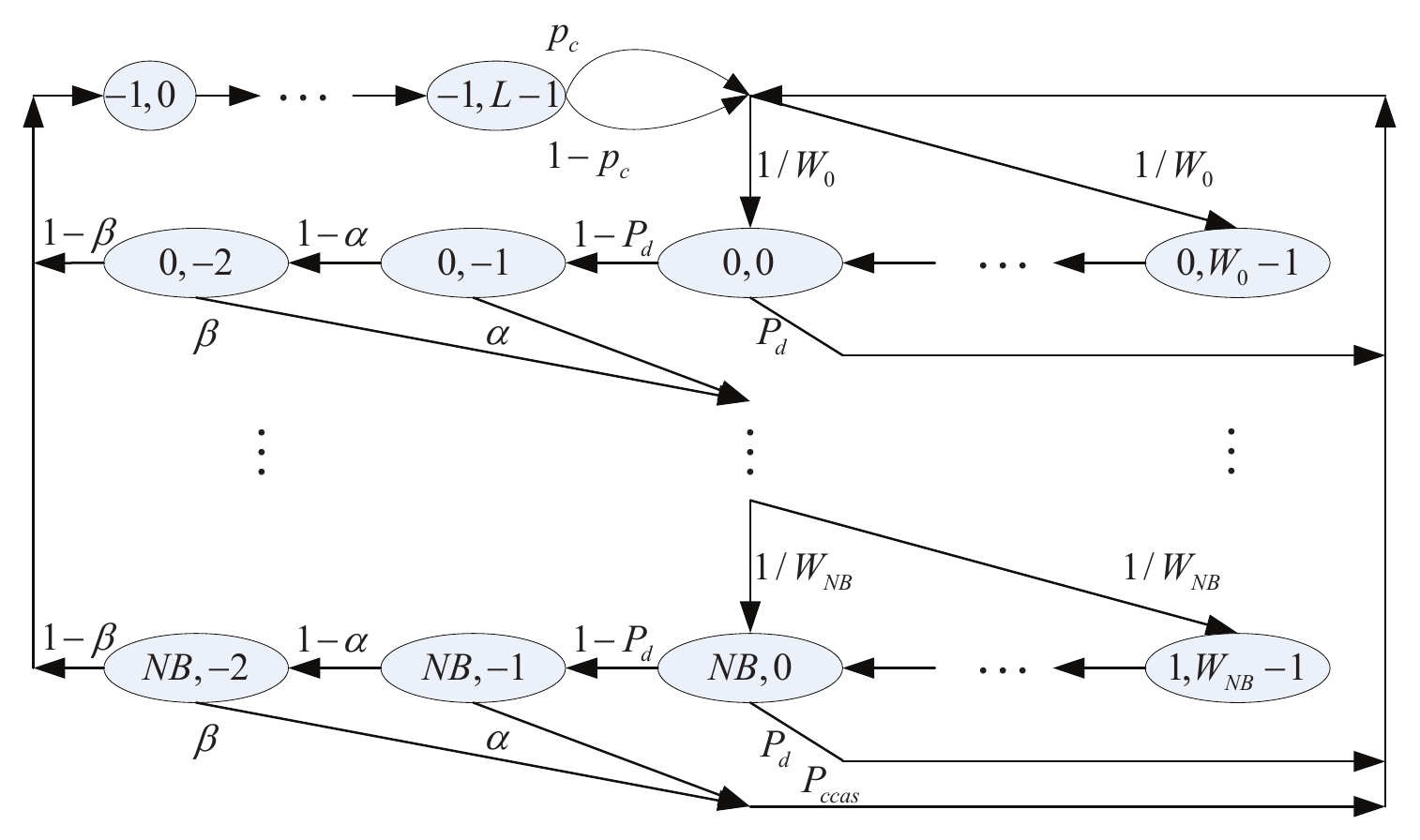} 
\caption{Markov chain model for CSMA protocol.}
\label{Markovchain}
\end{figure*}

We study the Markov chain model for the slotted CSMA/CA protocol, which is similar to the one in \cite{Prime}.
We consider the case with $h_i$ contending nodes and we use
the same notations as in \cite{Poll08}. However, our analysis has the additional ``deference state'',
which was not considered in \cite{Poll08}. 

We consider the 2D Markov chain (MC) for slotted CSMA/CA MAC protocol $\left(s(t), c(t) \right)$ where 
$s(t)=-1$ represents the transmission state, and $s(t)=[0,NB]$ captures the backoff stages (BK);
$c(t)=-1,-2$ denote the first and second CCAs; $c(t)=[0,L-1]$ represents transmission slots and $c(t)=[0,W_i-1]$
describes the states of backoff counter. 
The detail of superframe structure in one RI is demonstrated in Fig.~\ref{cycletime}.
Here, $L$ is used to denote the general transmission time
where $L=L_s = T_p + t_{ACK} + L_{ACK}$ for a successful transmission and $L=L_c = T_p + t_{ACK,ti}$
for a collided transmission, $T_p$ is packet transmission time, $L_{ACK}$ is the ACK time, and $t_{ACK,ti}$
is the ACK timeout. Fig.~\ref{Markovchain} shows the transition diagram of this MC. 

Let $b_{i,k} = {\sf lim}_{t \rightarrow \infty} \Pr\left(s(t) = i, c(t) = k\right)$ denote the stationary probability of the Markov chain.
In this paper, we asume that $NB = d$ as in \cite{Prime} where $d = macMaxBE - priority $, $priority = 2$ is the priority of node and also the number of CCAs.
In addition, $macMaxBE $ is the maximum backoff exponent and $NB = macMaxattempt-1 $ is the maximum number of backoffs.
We define the following parameters: $\lambda = \alpha + \beta - \alpha \beta $, $\omega = \lambda \left(1-\mathcal{P}_d\right) $, where $\alpha $ and $\beta $ are the probabilities that a node fails to identify an idle channel during $CCA_1 $ and $CCA_2 $, respectively; $\mathcal{P}_d = L_s/SF $ is the probability of deference. 

Using the analysis similar to that in \cite{Poll08}, we can arrive at the following the relationship for the steady-state probabilities: $b_{i,0} = \omega^i b_{0,0} $, $b_{i,-1} = b_{i,0} (1-\mathcal{P}_d) $, $b_{i,-2} = b_{i,0} (1-\mathcal{P}_d) (1-\alpha) $ (for $i \in \left[0, NB\right]$), and $b_{-1,k} = (1-\mathcal{P}_d) (1-\alpha) (1-\beta) \sum_{i=0}^{NB} b_{i,0} $, $b_{i,k} = \frac{W_i-k}{W_i} b_{i,0} $ (for $i \in \left[0, NB\right],  k \in \left[-2, {\sf max} \left(W_i-1, L_s-1\right)\right] $).
Since we have $\sum_{i,k} b_{i,k} = 1 $, substitute the above results for all $b_{i,k}$ and perform some manipulations, we can obtain
\beqn
\label{Stead_EQN}
1=\frac{b_{0,0}}{2} \left\{W_0 \frac{1-\left(2\omega\right)^{NB+1}}{1-2\omega}+ \frac{1-\omega^{NB+1}}{1-\omega} \times \right. \nonumber \\
\left. \left[3+2\left(1-\mathcal{P}_d\right)\left(1+\left(1-\alpha\right)\left(1+\left(1-\beta\right)L_s\right)\right)\right] \right\}. \label{b00_EQN}
\eeqn
From these results, we can find the relationship among $b_{0,0} $, $\alpha $, $\beta $, $\phi $ as follows \cite{Poll08}:
\beqn
\phi = \sum_{i=0}^{NB} b_{i,0} = \frac{1-\omega^{NB+1}}{1-\omega} b_{0,0} \label{phi1_EQN} \\
\phi = 1 - \left(1-\frac{\alpha}{L^*\left(1-\omega\right)}\right)^{\frac{1}{h-1}} \label{phi2_EQN} \\
\phi = 1 - \left(1 - \frac{\beta_{ACK}}{\left(1-\beta_{ACK}\right)\left(2-\mathcal{P}_{\text{ncol}}\right)}\right)^{1/h} \label{phi3_EQN}
\eeqn
where $\phi $ is the probability that a node is at the $CCA_1 $ state after backoff, $L^* = T_p + L_{ACK} \left(1 - \mathcal{P}_{\text{ncol}}\right)$, $\mathcal{P}_{\text{ncol}} = 1 - \frac{h\phi\left(1-\phi \right)^{h-1}}{1-\left(1-\phi\right)^h} $, $\beta_{ACK} = \frac{2-\mathcal{P}_{\text{ncol}}}{2-\mathcal{P}_{\text{\text{ncol}}}+\frac{1}{1-\left(1-\phi\right)^h}} $ .
From (\ref{b00_EQN}), (\ref{phi1_EQN}), (\ref{phi2_EQN}) and (\ref{phi3_EQN}), we can determine $b_{0,0} $, $\alpha $, $\beta $, and $\phi $ by using the standard numerical method.

\section{Calculation of ${\bar T}_i $ and $\sigma^2_i $}
\label{average_variance} 

In this appendix, we determine $\bar{T}_i $ and $\sigma^2_i $ for a particular SF $i$.
For simplicity, we again omit the index $i$ and $h_i$ in all related parameters when this does not create confusion.
First, we can express the probability generating function (PGF) of the generic frame, $T(z) $ which includes success, collision, CCA failure and deference,
as
\beqn
T\left(z\right) = \mathcal{P}_{\sf succ} T_S\left(z\right) + \mathcal{P}_{\text{coll}} T_C\left(z\right) + \mathcal{P}_{\text{ccas}} T_F\left(z\right) + \mathcal{P}_d T_D\left(z\right)
\eeqn
Here we denote $\mathcal{P}_{\text{ccas}} $, $\mathcal{P}_{\text{coll}} $ and $\mathcal{P}_{\text{succ}} $ as the probabilities of CCA failure, collision and success, respectively. These probabilities can
be calculated as $\mathcal{P}_{\text{ccas}} = \left(1-\mathcal{P}_d\right) \lambda^{NB+1} $, $\mathcal{P}_{\text{coll}} = p_c \left(1-\mathcal{P}_d\right) \left(1-\lambda^{NB+1}\right) $, and $\mathcal{P}_{\sf succ} = 1-\mathcal{P}_{\text{coll}}- \mathcal{P}_{\text{ccas}}- \mathcal{P}_d $, where $p_c = 1-\left(1-\phi\right)^{h-1} $.
Moreover, we also denote $T_S\left(z\right) $, $T_C\left(z\right) $, $T_F\left(z\right)$ and $T_D\left(z\right) $ as the PGFs of durations of success, collision, CCA failure and deference, respectively. 
These quantities can be calculated as in \cite{Park10}.

Finally, we can determine $\bar T $ and $\sigma^2$ from the first and second derivation of $T\left(z\right) $ at $z = 1 $, i.e.,
\beqn
\bar T = \frac{dT}{dz}\left(1\right) ; 
\sigma^2 = \frac{d^2T}{dz^2}\left(1\right) + \bar T - \left(\bar T\right)^2. \label{var_genfram_EQN}
\eeqn
These parameters $\bar T $ and $\sigma^2$ will be utilized in (\ref{K_genfram_EQN}).

\section{Calculation of ${E}_i $}
\label{energy_consumption_i} 

The energy consumption per node in SF $i$ with $h_i $ contending nodes can be expressed as 
\beqn
E_i = E_{b,i} + E_{{\sf ccas},i} + E_{{\sf suco},i} + E_{d,i}
\eeqn
where $E_{b,i}$, $E_{{\sf ccas},i}$, $E_{{\sf suco},i}$, and $E_{d,i} $ are the average energy consumption due to backoffs, CCAs, transmissions (successful/collided transmissions), and deference, respectively.
Let us denote $E_{\sf idle} $, $E_{\sf sens} $, $E_{\sf tx} $ and $E_{\sf rx} $ as the energy consumption
 corresponding to idle, CCA sensing, transmitting and receiving slots, respectively; then, these quantities are determined as follows.
For brevity, we again omit the index $i$ and $h_i$ in all related parameters if this does not create confusion.
First, the energy consumption during backoff duration is $E_{b,i} = E_{\sf idle} \sum_{j=0}^{NB} \sum_{k=0}^{W_j -1} b_{j,k} $.
After some simple manipulations, we can arrive at \cite{Park11}
\beqn
E_{b,i} = E_{\sf idle} /2 \left(W_0 b_{0,0} \frac{1-(2\omega)^{NB+1}}{1-2\omega}+3 \phi \right).
\eeqn
Also, $E_{{\sf ccas},i} $ can be determined as 
\beqn
E_{ccas,i} \!\!=\!\! E_{\sf sens}\!\sum_{j=0}^{NB} (b_{j,-1} \!+\! b_{j,-2}) \!\!=\!\! E_{\sf sens} (1-\mathcal{P}_d)(2-\alpha) \phi.
\eeqn
Moreover, $E_{{\sf suco},i} $ is given as \cite{Park11}
\beqn
E_{{\sf suco},i} = E_{\sf tx} \sum_{k=0}^{T_p-1} b_{-1,k} + E_{\sf rx} (1-p_c) \sum_{k=T_p+t_{ACK}}^{L_s-1} b_{-1,k} + \nonumber \\
E_{\sf idle} \left((1-p_c)\!\!\!\sum_{k=T_p}^{T_p+t_{ACK}-1} \!\! b_{-1,k}+p_c \!\! \sum_{k=T_p}^{T_p+t_{ACK,ti}-1} \!\! b_{-1,k}\right) \\
= (1-\lambda) (1-\mathcal{P}_d) \phi \left(E_{\sf tx} T_p + E_{\sf rx} L_{ACK} (1-p_c) + \right. \nonumber \\
\left. E_{\sf idle} (t_{ACK} (1-p_c) + t_{ACK,ti} p_c)\right).
\eeqn
Finally, $E_{d,i} $ can be expressed as $E_{d,i} = E_{\sf idle} \mathcal{P}_d L_s$.

\bibliographystyle{IEEEtran}


\begin{biography}[{\includegraphics[width=1in,height=1.25in,clip,keepaspectratio]{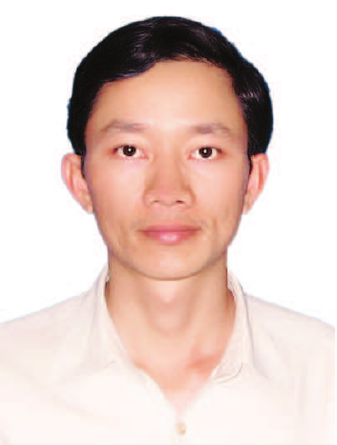}}]
{Le Thanh Tan} (S'11--M'15)  
 received the B.Eng. and M.Eng. degrees from Ho Chi Minh City University of Technology in 2002 and 2004, respectively and the Ph.D. degree from Institut National de la Recherche Scientifique--\'{E}nergie, Mat\'{e}riaux et T\'{e}l\'{e}communications (INRS--EMT), Canada in 2015. 
From 2002 to 2010, he was a Lecturer with the Ho Chi Minh University of Technical Education (HCMUTE). In 2015, he was a Postdoctoral Research Associate at \'{E}cole Polytechnique de Montr\'{e}al, Canada. He currently is a Postdoctoral Researcher at the School of Electrical Computer and Energy Engineering at Arizona State University (ASU), USA.
His current research activities focus on internet of things (IOT over LTE/LTE--A network, cyber--physical systems, big data, distributed sensing and control), time series analysis and dynamic factor models (stationary and non--stationary), wireless communications and networking,  Cloud--RAN, cognitive radios (software defined radio architectures, protocol design, spectrum sensing, detection, and estimation), statistical signal processing, random matrix theory, compressed sensing, and compressed sampling. He has served on TPCs of different international conferences including IEEE CROWNCOM, VTC, PIMRC, etc. He is a Member of the IEEE.
\end{biography}

\begin{biography}[{\includegraphics[width=1in,height=1.25in,clip,keepaspectratio]{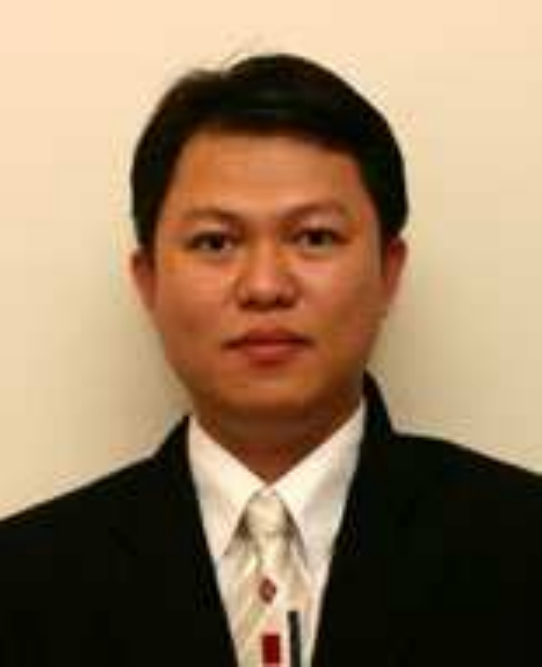}}]
{Long Le} (S'04--M'07--SM'12)  
received  the B.Eng.  degree  in  Electrical  Engineering  from  Ho Chi  Minh  City  University  of  Technology,  Vietnam, in 1999, the M.Eng. degree in Telecommunications from  Asian  Institute  of  Technology,  Thailand,  in 2002, and the Ph.D. degree in Electrical Engineering from  the  University  of  Manitoba,  Canada,  in  2007. He was a Postdoctoral Researcher at Massachusetts Institute  of  Technology  (2008--2010)  and  University  of  Waterloo  (2007--2008).  Since  2010,  he  has been  with  the  Institut  National  de  la  Recherche Scientifique  (INRS),  Universit\'{e} du Qu\'{e}bec, Montr\'{e}al,  QC,  Canada  where he  is  currently  an  associate  professor.  His  current  research  interests  include smart grids, cognitive radio, radio resource management, network control and optimization, and emerging enabling technologies for 5G wireless systems. He is a co-author of the book Radio Resource Management in Multi-Tier Cellular Wireless Networks (Wiley, 2013). Dr. Le is a member of the editorial board of  IEEE  TRANSACTIONS  ON  WIRELESS  COMMUNICATIONS,  IEEE COMMUNICATIONS SURVEYS AND TUTORIALS, and IEEE WIRELESS COMMUNICATIONS LETTERS. He has served as a technical program committee  chair/co-chair  for  several  IEEE  conferences  including  IEEE  WCNC, IEEE VTC, and IEEE PIMRC.
\end{biography}
\end{document}